\documentstyle[aps,12pt,psfig, subfigure,preprint]{revtex}
\tightenlines
\newcommand{\be}{\begin{equation}}
\newcommand{\ee}{\end{equation}}
\newcommand{\ba}{\begin{eqnarray}}
\newcommand{\ea}{\end{eqnarray}}
\newcommand{\nn}{\nonumber}

\newcommand{\ti}[1]{\bar{#1}}
\newcommand{\la}{\langle}
\newcommand{\ra}{\rangle}  
\newcommand{\di}{\displaystyle}

\begin{document}

\newpage
\vspace*{2cm}

\begin{center} 
{\LARGE {\bf Nonlinear reactive systems on lattice viewed as Boolean dynamical
systems} }
\vskip2cm
E. Abad\footnote[1]{E-mail: eabad@ulb.ac.be }, 
 P. Grosfils and G. Nicolis\\  
\vspace*{0.2cm}
Centre for Nonlinear Phenomena and Complex Systems\\
Universit\'e Libre de Bruxelles\\
Campus Plaine C.P. 231\\
B-1050 Bruxelles 
\\
\vspace*{0.5cm}
\today\\
\end{center}
{\bf Abstract}
We present a stochastic, time-discrete boolean model which mimics the 
mesoscopic dynamics of the desorption reactions $A+A\rightarrow A+S$ and 
$A+A\rightarrow S+S$ in a 1D lattice. In the continuous-time limit, 
we derive a hierarchy of dynamical equations for the 
subset of moments involving contiguous lattice sites. The solution 
of the hierarchy allows to compute the exact dynamics 
of the mean coverage for both microscopic and coarse-grained initial 
conditions, which turn out to be different from the mean field predictions. 
The evolution
equations for the mean coverage and the second order moments are 
shown to be equivalent to those provided by a time-continuous 
Master equation. The important role of higher order fluctuations 
is brought out by the failure of a truncation scheme retaining only
two-particle fluctuation correlations.\\[.5cm]
PACS number(s): 05.40.-a, 05.45.-a\\
\section{Introduction}
When reactive processes take place in low dimensional spaces, 
the mean field (MF) laws of classical kinetics, in which each particle is 
assumed to interact with the system as a whole, become questionable.
In this simple one-particle picture, information about 
special geometric constraints or
 many-particle correlation effects is absent. However, these features become 
increasingly important as the dimensionality of the space decreases and may
eventually give rise to a crossover between universal macroscopic MF behavior
 and specific, strongly lattice-dependent behavior below a critical
dimension $d_c$. The critical dimension $d_c$ is found to depend 
on the characteristics of the support (such as average coordination number) 
and the stoichiometry of the reactive dynamics, i.e., its degree of 
nonlinearity. 
 
The crossover phenomenon mentioned above is not specific of reactive systems, 
but is common to many other statistical systems 
(interacting spins, random walks, etc.).
 To understand its real nature, it is necessary to 
resort to a more refined description in which relevant fluctuation
effects in concentration and occupation number space can be accounted for. 
In view of the difficulties encountered by a full-scale microscopic analysis, 
one typically tries to devise 
``minimal'' models which contain the necessary ingredients to observe 
a significant departure of MF behavior at sufficiently low dimensions.
A probabilistic, time-continuous approach of this kind for a detailed 
study of fluctuations is provided by Master equation (ME) techniques, 
in which the local dynamics is
specified by transition rates between different microscopic states. In 
the ME approach, one is typically 
led to a set of evolution equations for the moments of the probability 
density, which can be taken as a starting point to compute 
the most characteristic
macroscopic observables, generally associated with the first low-order 
moments. However, if the transition rates are nonlinear, the
low-order moments will not obey closed equations but will rather be
linked to the higher order ones by an infinite hierarchy of coupled 
equations. In the general case, these hierarchies cannot be solved 
in a closed form. Suitable truncation techniques may 
then help to overcome this difficulty. 
  
 A more recent, very successful method to study the evolution of low 
dimensional reactive systems is based on Monte Carlo (MC) simulations
\cite{Alba}, in which the microscopic state of 
the system is updated at discrete time steps according to a given 
dynamical rule. One of our goals in this 
paper is to stablish a link between the physics underlying 
this simulational approach
and the ME formalism. To this end, we shall formulate mathematically the 
evolution law prescribed
by the MC algorithm as a stochastic dynamical system whose state variables
are  Boolean 
occupation numbers for the state of each lattice site \cite{Boon,Thom}. 
As we shall see, this 
cellular automaton (CA) model yields a set of moment equations similar to 
that obtained from a time-continuous ME.  

 So far, most of the literature devoted to the effects of dimensionality
on reactive dynamics has focused on diffusion-controlled reactions 
\cite{Lig,Mikh,Glas}. In these systems, deviations of MF behavior at low 
dimensions may be expected due to the reduced effective mobility of 
the reactants. There are, however, situations in which diffusion can 
be neglected within the time scale of interest, like e.g.
chemical reactions in solid materials and certain radical
isolation problems \cite{Coh}. Such systems of immobile reactants constitute
the object of our study in this paper. Clearly, the absence of diffusion 
reduces the ability 
of each particle to interact with all the others, thus driving
the system further away from the applicability conditions of MF theory. This
effect is enhanced in the presence of short range 
interactions restricted to, say, nearest neighbours and hard core exclusion 
not allowing more than one particle per lattice site 
\cite{Schn1,Schn2,Priv,Prov1,Prak}. 

More specifically, we will study the 1D lattice dynamics in two particular 
examples of nonlinear irreversible associated with 
the cooperative desorption systems (CDs)
\begin{mathletters} 
\ba
\label{pade}
A+A & \stackrel{k_R}{\longrightarrow} & A+S \qquad 
\mbox{cooperative partial desorption (CPD)} \\
\label{rede}
A+A & \stackrel{k_R}{\longrightarrow} & S+S.\qquad 
\mbox{cooperative total desorption (CTD).} 
\ea 
\end{mathletters}
where $A$ is the reactive species, $S$ the empty lattice site and $k_R$
is the rate of reaction. Notice that steps (\ref{pade})-(\ref{rede}) are 
typical parts of 
realistic, more complex reaction schemes as discussed further in the 
Conclusions.
Despite their apparent simplicity, they display a complex non-MF 
behavior characterized by frozen, non universal, initial-condition-dependent 
steady states \cite{Priv,Ken,Fot}. The dependence on the initial conditions 
is the signature of the weak ergodic properties of the CDs, 
which is in turn related to the irreversibility of the reactive 
schemes (\ref{pade})-(\ref{rede}). A remarkable
feature of the CDs is the special structure of the underlying moment 
equations, which
allows one to compute explicitly the coverage and the fluctuation dynamics
and thus test in a very efficient way the applicability of diverse truncation 
approaches.     

The paper is organized as follows. In section \ref{cpd}, a comprehensive
study of the one-dimensional CPD is presented. First, we introduce a 
particular biased implementation of the CPD and 
characterize its non-MF behavior using
an intuitive argument. In \ref{bomo}, we introduce a stochastic
dynamical rule which mimics the microscopic, time-discrete evolution of 
the system as prescribed by MC simulations and show 
that this dynamical
rule characterizes properly the steady states of the system. In 
\ref{moeq}, we derive evolution equations for the first-
and second-order moments of the probability distribution for the 
microscopic states and use the same formalism to obtain a  
hierarchy of coupled equations describing the time evolution of a particular
subset of moments, namely those involving
contiguous occupation numbers. These moments can be identified
with the probabilities of finding a randomly chosen string of contiguous
sites simultaneously occupied. In \ref{meco}, it is found that 
the solution of the hierarchy 
for both microscopic and coarse-grained initial conditions leads to an 
explicit expression for the mean coverage of the lattice. We also derive
equations for the coarse-grained dynamics of vacant sites in terms of 
generalized probabilities for occupied sites.  
Next, in \ref{ME}, we present a general ME for an array of 
Boolean variables and show that, in the special case of the 
CPD, the evolution equations for 
the mean occupation number and the second-order moment are similar to those 
obtained in the CA approach. 
In section \ref{trsc}, we study the relevance of inhomogeneous 
fluctuations in the CPD system. A set of evolution equations retaining 
only pair fluctuation
correlations fails to reproduce the correct dynamics, the reason being
the nonnegligibility of higher order fluctuation correlations. These can be 
calculated explicitly from sets of generalized moment equations, which also
allow us to compute the dynamics of the vacant sites for clusters of small
size.    
In section \ref{ctd}, we extend the main 
results of sections \ref{cpd} and \ref{trsc} to the CTD. The
conclusions are summarized in section \ref{suol}, which also outlines a 
research strategy for future work in this area. In the Appendices,
we deal with a symmetric version of the CPD and discuss its equivalence 
with the biased CPD. We also study the dynamics of 
particle islands.                      
\section{ Boolean dynamics of the CPD: Model and exact results}
\label{cpd}
\subsection{Definition of the system and MF rate equation}
\label{defsys}
Consider a 1D lattice with $N$ sites which may be either empty ($S$) 
or filled with a single particle ($A$). The reaction takes place 
according to the scheme of Fig. \ref{pdstep}. At each time step $\Delta t$, 
a lattice site is randomly 
chosen. If the selected site is filled, the particle desorbs with 
a certain probability (equal to the rate of reaction $k_R$) provided 
that its right neighbour site is also occupied; otherwise, 
nothing occurs. 
The classical rate equation for this CPD scheme (Eq. (\ref{pade})) reads
\be
\label{mfeq}
\frac{d c(t)}{dt}=-k_R \, c(t)^2,  
\ee
where $c(t)$ is the A particle concentration, here identified as 
the mean fraction of occupied sites, also referred as ``coverage''. 
Its solution predicts a decay of the form 
\be
c(t)=\frac{c(0)}{1+c(0)\, k_R\, t}.
\ee
to a zero-coverage steady state. However, as we see below, the restricted 
1D geometry of the lattice allows for non MF 
steady states. It is indeed easily seen that all lattice configurations
without contiguous particles cannot evolve in time. To understand this 
obvious MF failure in detail, we need to set up a model for the 
microscopic dynamics.      
\subsection{Boolean modeling and steady states}
\label{bomo}
We stipulate that each site in the lattice is characterized by a Boolean 
occupation number:
\be
n_i(t)= \left \{ { \quad 1 \qquad \mbox{if site i is occupied} \atop
0 \qquad  \mbox{if site i is empty} }, \qquad i=1,\ldots,N. \right.
\ee 
Each site evolves according to the nonlinear dynamical rule 
\be
\label{dynrul}
n_i(t+\Delta t)=n_i(t)-\xi_N^{(i)}(t)\, \xi_R(t)\,n_i(t)\,n_{i+1}(t), 
\quad i=1,\ldots, N 
\ee  
where 
\be
\vec{\xi}_N(t)=\left( \begin{array}{c}
\xi_N^{(1)}(t) \\ \vdots \\ \xi_N^{(N)}(t)
\end{array} \right), \qquad \xi_N^{(i)}(t)=0,1.
\ee 
is a $N$-dimensional vector of stochastic Boolean decision variables. If, 
say, site $i$ is chosen, the $i$-th component will be one and all other
components will be zero. Hence, the different components $\xi_N^{(i)}(t)$ 
in a given, single realization are correlated:
\be
\label{corstp}
\sum_{i=1}^N \xi_N^{(i)}(t)=1\qquad \qquad
\xi_N^{(i)}(t)\, \xi_N^{(j)}(t)=\xi_N^{(i)}(t)\, \delta_{i j}.
\ee
Since at each time step $\Delta t$, one site out of $N$ is randomly chosen, 
the mean value $\overline{\xi_N^{(i)}(t)}$ over 
an ensemble of realizations will be $1/N$. The additional stochastic 
variable $\xi_R(t)$ takes randomly the values $1$ 
and $0$ with probabilities $p_R$ and $1-p_R$, respectively. 

To have a well posed problem, we still have to specify the boundary 
conditions. We shall take periodic boundary conditions, which amounts to 
setting $n_{N+1}(t)=n_1(t)$ in the last equation (\ref{dynrul}).

Implicit in the above scheme is the idea that, at most, 
one reactive event may occur in the lattice in a given time interval 
$\Delta t$, even if there are several 
a priori reactive pairs. The justification of such a prescription is that, 
typically, the ocurrence of a reaction requires to overcome an activation 
energy threshold. At ordinary temperatures
(supposed to prevail here), this can be achieved only if a sufficiently 
strong fluctuation impinges on the system as a result, say, of the coupling 
between the adsorbate and the lattice. Since the probability of such an 
event is small, one may expect that reactions will occur asynchronously 
at different lattice sites, which is precisely what is stipulated in 
our evolution rule.    

Eqs. (\ref{dynrul}) will be used in the sequel to describe the 
dynamics of the CPD at three different levels: 
\begin{enumerate}
\item A ``microscopic'' level in which the exact microscopic 
dynamics, i.e. the initial condition $\{n_i(0)\}$ and the decision path 
represented by the variable sets $\{\vec{\xi}_N(0),\ldots,\vec{\xi}_N(t)\}$ 
and 
$\{\xi_R(0), \ldots \xi_R(t) \}$, is assumed to be known in detail. 
\item an intermediate level in which the system is prepared in such a way, 
that the initial condition is 
known in detail and one averages only over the subsequent dynamics, i.e. ,
over the different paths or realizations. 
\item a {``coarse-grained''} description in which one averages over a
statistical ensemble of realizations {\it and} a nonequilibrium
ensemble of initial conditions.
\end{enumerate}
After a time $t_{st}$ a steady state $\{n_i^{st}\}$ is attained. 
The dynamical rule (\ref{dynrul}) allows a straightforward characterization
of these steady states. Since the first term in the r.h.s. of (\ref{dynrul})
cancels with the l.h.s., one has:
\be
\xi_N^{(i)}(t)\,n_i^{st}\,n_{i+1}^{st}=0.
\ee
This holds for all times $t\ge t_{st}$. Since $\xi_N^{(i)}(t)$ is a random 
variable, it will be nonzero at some time $t$ for any given site $i$,
meaning that the product $n_i^{st}\,n_{i+1}^{st}$ must vanish for all $i$. 
This expresses the 
fact that two contiguous sites cannot be simultaneously occupied in a 
steady state. However, not all configurations satisfying this condition
are attained with equal probability. For instance, if $N$ is even, we can be
certain that the 
steady-state configuration with alternating occupied and empty sites 
will never be reached if we start with less than half the lattice filled.
Therefore, a simple combinatorial counting of the non-evolving configurations
 fails to provide the correct mean steady-state coverage \cite{Fot}.
\subsection{Moment equations and cluster dynamics}
\label{moeq}
 Let us now turn to the dynamics generated by Eqs. (\ref{dynrul}). First, 
we fix the initial 
configuration $(n_1(0),\ldots, n_N(0))$ of the lattice and take 
the average of Eq. (\ref{dynrul}) over an ensemble 
of different realizations. Using the statistical independence of 
$\xi_N(t)$, $\xi_R(t)$ and the occupation numbers $n_i(t)$, this yields
\be
\label{avevol}
\overline{n_i(t+\Delta t)}-\overline{n_i(t)}=
-\,\frac{p_R}{N}\,\overline{n_i(t)\, n_{i+1}(t)}, 
\qquad i=1,\ldots, N.
\ee
The quantity $\frac{\di p_R}{\di N}$ can be regarded as the probability 
that a reaction 
takes place at a given site $i$ in the time interval $(t,t+\Delta t)$
given that site $i$ and $i+1$ are occupied. The rate of reaction is 
obtained by dividing this probability by the time step $\Delta t$:
\be
k_R=\frac{p_R}{N\Delta t}
\ee   
To derive the continuous-time limit of Eq. (\ref{avevol}), we divide 
both sides by $\Delta t$ and let this quantity go to zero in such a way that 
the reaction rate $k_R$ remains finite. This can be e.g. accomplished
by letting the system size $N$ simultaneously go to infinity so 
that $N\Delta t=C$. We shall choose the constant $C$ equal to unity, 
implying that, after one time unit, each lattice site has been visited
once on average. With this choice, $k_R$ becomes numerically equal
to the value of $p_R$. 
The resulting equation reads     
\be
\label{meanoc}
\frac{d}{dt}\overline{n_i(t)}=-k_R\,\overline{n_i(t) n_{i+1}(t)}
\ee
In a similar way, one can obtain an evolution equation for the 
second-order moment 
$\overline{n_i(t) n_j(t)}$. Multiplying (\ref{dynrul}) 
by $n_j(t+\Delta t)$ and using (\ref{corstp}), one finds
\be
\label{discor2}
\overline{n_i(t\!+\!\Delta t)\,n_j(t\!+\!\Delta t)}\!-\!
\overline{n_i(t)n_j(t)}
=
-\frac{p_R}{N}\,\overline{n_i(t) n_{i+1}(t) n_j(t)}
-\frac{p_R}{N}\,\overline{n_i(t) n_j(t) n_{j+1}(t)},  
\quad i\neq j
\ee 
which in the continuous-time limit becomes
\be
\label{cor2}
\frac{d}{dt}\,\overline{n_i(t) n_j(t)}=
-k_R\,\overline{n_i(t) n_{i+1}(t) n_j(t)}-k_R\,
\overline{n_i(t) n_j(t) n_{j+1}(t)}.
\ee
For $j=i+1$, Eq. (\ref{cor2}) together with the Boolean property 
$n_i^2(t)=n_i(t)$ yields 
\be 
\label{cl2}
\frac{d}{dt}\,\overline{n_i(t) n_{i+1}(t)}=
-k_R\,\overline{n_i(t) n_{i+1}(t)}-
k_R\,\overline{n_i(t) n_{i+1}(t) n_{i+2}(t)}
\ee
Unless otherwise specified, we shall set $p_R=1$ for simplicity, implying that
$k_R$ is equal to unity.
Note, however, that it is possible to 
absorb the reaction rate $k_R$ appearing in the time-continuous equations 
(\ref{meanoc}) and (\ref{cor2}) in the time scale by means of a 
dimensionless variable $\tau=k_R\,t$ \cite{Priv,Fot}. 

An important feature of Eqs. (\ref{meanoc}) and (\ref{cl2}) is that they
couple in a linear fashion moments involving only contiguous sites.    
This turns out to be a generic property that also holds for the 
dynamics of higher order moments. 

Let us now consider a more general subset of $k$ contiguous lattice sites 
$i,\ldots,i+k-1$. 
The probability to find all sites occupied at time $t$, i.e., a cluster
of size $k$, is given by the quantity
\be
M^{(i)}_k(t)=\overline{\prod_{j=i}^{i+k-1}n_j(t)}, \qquad i,k=1,\ldots,N.
\ee
To derive an evolution equation for $M^{(i)}_k(t)$, we must generalize Eqs.
(\ref{meanoc}) and (\ref{cl2}). Taking the dynamical rule (\ref{dynrul})
as a starting point, one obtains
\be
\label{prodev}
\prod_{j=i}^{i+k-1}n_j(t+\Delta t)-\!\!\prod_{j=i}^{i+k-1}n_j(t)=
-\!\!\sum_{j=i}^{i+k-2}\xi_N^{(j)}(t)\,\xi_R(t)\prod_{j=i}^{i+k-1}n_j(t) 
-\xi_N^{(i+k-1)}(t)\,\xi_R(t)\prod_{j=i}^{i+k}n_j(t),
\ee
where $i=1,\ldots,N $. Taking averages in (\ref{prodev}) for $k<N$,
one gets
\ba
\label{dismo}
M^{(i)}_k(t+\Delta t)-M^{(i)}_k(t)=-\frac{(k-1)}{N}\,M^{(i)}_k(t)-
\frac{1}{N}\,M^{(i)}_{k+1}(t).  
\ea
We now introduce the global quantity   
\be
\ti{P}_k^{(N)}(t)=\frac{1}{N}\sum_{i=1}^N M^{(i)}_k(t),
\ee
representing the probability of 
finding $k$ filled adjacent sites in the $N$-site lattice. 
Summing over the site index $i$ in  Eqs. (\ref{dismo}), we find
\be
\label{dispriv}
\ti{P}_k^{(N)}(t+\Delta t)-\ti{P}_k^{(N)}(t)=-\frac{(k-1)}{N}\,
\ti{P}_k^{(N)}(t)-\frac{1}{N}\,\ti{P}_{k+1}^{(N)}(t), \quad k<N.
\ee 
Taking the continuous-time limit of 
(\ref{dispriv}) for a given finite value of $k$,
a set of evolution equations for $\ti{P}_k(t)=
\lim_{N\to\infty}\ti{P}_k^{(N)}(t)$ is obtained: 
\be
\label{privpone}
\frac{d\ti{P}_k(t)}{dt}=-(k-1)\,\ti{P}_k(t)-\ti{P}_{k+1}(t), 
\qquad k=1,2,\ldots,k_{max},
\ee
where the integer $k_{max}$ is the size of the largest cluster. The first
term on the r.h.s. of (\ref{privpone}) corresponds to a destruction of a
k-particle cluster by the interaction of two particles inside the cluster, 
while the second term stands for its destruction by desorption of its 
rightmost particle, which requires an additional occupied site, i.e., 
a cluster of $k+1$ sites. 
 
If the lattice contains only clusters 
of finite size, then $k_{max}<\infty$ and the hierarchy of equations 
(\ref{privpone}) is truncated by the condition 
$\ti{P}_{k_{max}+1}(t)\equiv 0$. The last equation (\ref{privpone}) 
reads then
\be
\label{lapriv}
\frac{d\ti{P}_{k_{max}}(t)}{dt}=-(k_{max}-1)\,\ti{P}_{k_{max}}(t).
\ee
\subsection{Solution of hierarchy and mean coverage 
for microscopic and coarse-grained initial conditions}
\label{meco}
Suppose that we start with a given initial configuration 
$\left\{n_i(0)\right\}$ characterized by the nonvanishing set of 
probabilities
\be
\label{init}
\left \{ \ti{P}_1(0),\ti{P}_2(0),\ldots,\ti{P}_{k_{max}}(0) \right \}. 
\ee
Using the linearity of 
Eqs. (\ref{privpone}), one can show that the solution for $t>0$ is
of the form
\be
\label{gensol}
\ti{P}_k(t)=\exp(-(k-1)t)\sum_{j=0}^{k_{max}-k} 
\frac{(\exp(-t)-1)^j}{j!}\ti{P}_{k+j}(0), \quad k=1,\ldots,k_{max}. 
\ee  
Note that for $t\to \infty$ only $\ti{P}_1$ survives, representing the
probability that a randomly chosen site be occupied. Clearly, this 
probability is 
equal to the ratio between the average number of occupied sites and 
the total number of sites, i.e., the mean coverage of the lattice
$c(t)$. 
In the long-time limit, one has:
\be
\label{genasco}
c(\infty)=\ti{P}_1(\infty)=\lim_{N\to\infty}\frac{\overline{N_A(\infty)}}{N}=
\displaystyle{\sum_{j=0}^{k_{max}-1}}
\,\frac{\textstyle (-1)^j}{\textstyle j!}\, \ti{P}_{j+1}(0).
\ee
We see that the asymptotic value of the coverage depends on all
details of the initial cluster distribution.   
The validity of Eq. (\ref{genasco}) is confirmed by MC simulations 
(Fig. \ref{pdcov}(a) ). The simulations were performed 
according to the time-discrete rule (\ref{dynrul}). The data correspond to a 
chain ring of $100$ particles and the initial configurations:
\begin{eqnarray*}
C1&=& 11111111111111110111111111111011110111111111111111011111111111111111\\
&&11101111111011011111111011111010 \\
C2&=& 01111001111110010011110111110111011111111110100011111111011011001111\\
&&11011101001111101011111111101111\\
C3&=& 11010101101001100100101010010010000101001011101001001001011001001111\\
&&01100011111001011001101001100111\\
C4&=& 10011100111000010011011000001000000000010000100100100100011000101000\\
&&01001010010101000011001101100010
\end{eqnarray*}

Let us now assume the more realistic situation in which the 
initial configuration of the system is not known in detail, but is 
characterized only through the particle coverage $p$. If we consider a 
uniform ensemble of all random microscopic configurations 
with a given $p$, the system will initially be translationally 
invariant and this property will 
propagate in time. Thus, the probabilities of finding $k$ consecutive 
filled sites $i,\ldots,i+k-1$ 
\be
P_k^{(N)}(t)=\langle \prod_{j=i}^{i+k-1} n_j(t) 
\,\rangle, \qquad k= 1,\ldots N
\ee
(the angular brackets $\la\cdots\ra$ stand for the averaging over
 realizations {\it and} initial states) will not depend on the evaluation 
site $i$. After averaging 
Eq. (\ref{dismo}) over the initial conditions and going over to the 
time continuum, one 
obtains:
\be
\label{cgpriv}
\frac{d P_k(t)}{dt}=-(k-1)\,P_k(t)-P_{k+1}(t) 
\qquad k=1,2,\ldots,
\ee
which is formally identical to Eq. (\ref{privpone}) 
for $\ti{P}_k$ with $k_{max}\rightarrow\infty$. However, the initial 
conditions are now simpler. Since the initial occupancies of each 
site are statistically independent, one has:
\be
\label{inco}
P_k(0)=p^k, \qquad k=1,2,\ldots
\ee
A solution satisfying the initial conditions (\ref{inco}) can be easily 
found by using an ansatz suggested in \cite{Priv}. One obtains
\begin{mathletters}
\label{dynpk}
\ba
\label{cgpro}
P_1(t)&=&p\, \exp{\,[ p\,(\exp{(-t)}-1)}\,], \\
\label{cgpro2}
P_k(t)&=&p^{k-1}\exp\left [-(k-1)t\right] P_1(t),\quad k=2,3,\ldots
\ea 
\end{mathletters}
Again, the survival expectancy $P_1(t)$ corresponds to the mean 
coarse-grained coverage 
$c(t)=\lim_{N\to\infty}\frac{\di \la N_A(t) \ra}{\di N}$. 
For short times, the series expansion of (\ref{cgpro}) yields:
\be
c(t)=P_1(t)=p-p^2\,t+\frac{p^2(p+1)}{2}\,t^2+o(t^3)
\ee
It is interesting to compare this with the short-time asymptotics of 
the exact solution of the MF equation (\ref{mfeq}) for $k_R=1$:
\be
c_{MF}(t)=\frac{p}{1+p\,t}=p-p^2\,t+p^3\,t^2+o(t^3) 
\ee
Thus, we see that the first MF deviation appears in the second order term. 
For long times, one observes an exponential decay into an absorbing state 
characterized by the asymptotic coverage 
\be
\label{asco}
c(\infty)=p\,\exp{(-p)} 
\ee
Formulae (\ref{dynpk}) for the transient behavior and 
(\ref{asco}) 
for the asymptotics are confirmed by time-discrete MC simulations
(Fig. \ref{pdcov}(b) ). As in the previous case, all 
$P_k(t)$ with $k\ge2$ vanish for $t\to\infty$.
Note that Eqs. (\ref{dynpk}) are formally obtained from 
Eqs. (\ref{gensol}) by 
 averaging over a uniform 
distribution of microscopic states with global coverage $p$. This amounts
to allowing for clusters of arbitrarily large size , i.e., 
$k_{max}\to\infty$.

So far, we have only considered the dynamics of particles, but it is also
of interest to
ask for the time evolution of empty intervals of a given size 
inside the lattice. Clearly, the number of empty intervals must grow 
as the reactions proceed. To set up explicit equations for clusters 
of vacant sites (``holes''), we consider the time-discrete 
evolution equation for the complementary occupation number 
$s_i(t)=1-n_i(t)$:
\be
s_i(t+\Delta t)-s_i(t)=\xi_N^{(i)} n_i(t)n_{i+1}(t),\qquad i=1,\ldots,N.
\ee
By averaging this equation over an ensemble of realizations and a 
uniform initial distribution of microscopic states with a 
coverage $p$ one gets:
\begin{mathletters}
\label{holeq}
\be
\label{hol1}
\frac{dS_1(t)}{dt}=P_2(t),
\ee
where $S_1(t)=\langle s_i(t)\rangle$. Eq. (\ref{hol1}) is in agreement 
with the conservation law $\frac{\displaystyle d}{\displaystyle dt}
(P_1(t)+S_1(t))=0$.

The dynamics of hole multimers can be derived starting from the
evolution equation for 
the product ${\displaystyle \prod_{k=i}^{i+k-1} s_k(t)}$. 
The general form of the evolution equation 
for $S_k(t)=\displaystyle{\langle \prod_{k=i}^{i+k-1} s_k(t)\rangle}$ 
becomes more and more complicated with increasing cluster size. 
The next three equations read  
\ba
\label{hol2}
\frac{dS_2(t)}{dt}&=&P_2(t)-P_3(t),  \\
\label{hol3}
\frac{dS_3(t)}{dt}&=&P_2(t)-P_3(t)+P_4(t)-P_{1,2}^{(2)}(t),  \\
\label{hol4}
\frac{dS_4(t)}{dt}&=&P_2(t)\!-\!P_3(t)\!+\!P_4(t)\!-\!P_5(t)\!-
\!P_{1,2}^{(2)}(t)\!+\!P_{1,3}^{(2)}(t)\!+\!P_{2,2}^{(2)}(t)\!-
\!P_{1,2}^{(3)}(t),
\ea
\end{mathletters}
where the quantities $P_{n,m}^{(l)}(t)$ on the r.h.s.  
of Eqs. (\ref{hol3})-(\ref{hol4}) are the joint probabilities of finding two 
fully occupied strings of length $n$ and $m$ separated by $l$ lattice 
spacings. The initial conditions for which these equations must be solved 
are
\be
S_k(0)=(1-p)^k, \qquad k=1,2,\ldots
\ee 
Eqs. (\ref{hol1}) and (\ref{hol2}) can be integrated straightforwardly, 
since the functions on the r.h.s. are already explicitly known. In section 
\ref{trsc}, we show how to obtain explicit expressions for the 
probabilities $S_3(t)$ and $S_4(t)$ by solving a generalized hierarchy for 
the joint probabilities $P_{n,m}^{(l)}(t)$.   
\subsection{ Comparison with the ME approach}
\label{ME}
An alternative way of deriving moment equations in the framework of a 
time-continuous approach is provided by the Master equation (ME).
Let us again consider a set of $N$ Boolean variables $\{n\}=(n_1,\ldots, n_N)$ 
arranged on the sites of a 1D lattice with periodic 
boundaries ($n_{N+1}=n_1$). 
Starting from an arbitrary state, the evolution of the 
probability density $P(\{n\}; t)$ is governed by a ME of the form:
\be
\label{meq}
\frac{d P(\{n\}; t)}{dt}=-\sum_j w_j(\{n\} \rightarrow \{n'\}, t)
 P(\{n\};t)+\sum_j w_j(\{n'\} \rightarrow 
\{n\}, t) P(\{n'\};t)
\ee
where $\{n'\}=(n_1,\ldots,1-n_j,\ldots,n_N)$ and 
$w_j(\{n\} \rightarrow \{n'\}, t)$ is the transition rate from the state
$\{n\}$ to the state $\{n'\}$ at time $t$. To obtain an evolution 
equation for the average occupation number  
\be
\langle\, n_i(t)\,\rangle=\sum_{\{n\}}n_i(t)\,P(\{n\},t),
\ee
we multiply Eq. (\ref{meq}) by $n_i(t)$ and sum over all 
different configurations $\{n\}$. This yields: 
\be
\label{memean}
\frac{d}{dt}\, \langle n_i(t) \rangle =\langle\, \left[1-2n_i(t)\right]\,
w_i(\{n\} \rightarrow \{n'\}, t)\, \rangle
\ee
For the time evolution of the second order moment one gets
\ba
\label{mecor2}
\frac{d}{dt}\, \la n_i(t) n_j(t) \ra &=&
\langle\,\left[n_j(t)-2n_i(t)\,n_j(t)\right]\, 
w_i(\{n\} \rightarrow \{n'\}, t)
 \,\rangle \nn \\
&& +\langle \,\left[ n_i(t)-2n_i(t)\,n_j(t)\right]\, 
w_j(\{n\} \rightarrow \{n'\}, t)\,\rangle
\ea
For the CPD, the transition rate reads  
\be
\label{astrpr}
w_j(\{n\} \rightarrow \{n'\}, t)=k_R\,n_i(t)\,n_{i+1}(t).
\ee
Using Eqs. (\ref{memean}) and (\ref{mecor2}) and the transition probability 
(\ref{astrpr}) with $k_R=1$, one can formally recover Eqs. (\ref{meanoc}) and 
(\ref{cor2}) averaged over an ensemble of random initial conditions. As
expected, this also holds for the higher order moments \cite{Fot}.  Note 
that in the ME approach the double averaging over realizations and initial 
conditions
is performed {\it simultaneously} via the probability distribution 
$P(\{n\},t)$. In the CA model, this has been done in two 
separate steps. One first chooses 
randomly an initial condition with a given $p$ and thereafter performs
a series of different realizations. Next, one again selects 
a new random initial condition and a new set of realizations is carried out, 
and so on.  
\section{Truncation schemes and role of fluctuations in the CPD}
\label{trsc}
The statistical properties of the N-site CPD system 
are described by the ``Boltzmann-like'' distribution functions
 $c_{i_1,i_2 \ldots ,i_n}^{(N)}(t)= \la n_{i_1}(t)\,n_{i_2}(t),
\ldots ,n_{i_n}(t)\ra $, which can be expressed in terms of  
the mean occupations $c_i^{(N)}(t)\,=\,\la n_i(t)\ra$ and a set of 
correlation functions
\be
\label{a1}
f_{i_1\,i_2,\dots,i_n}^{(N)}(t_1;\dots;t_n\,)\,=\,
\la\,\delta n_{i_1}(t_1)\,\delta n_{i_2}(t_2)\dots
\delta n_{i_n}(t_n)\,\ra,
\ee     
where $\delta n_i(t)\,=\,n_i(t)\,-\,c_i^{(N)}(t)$. In the following, 
we shall omit the superscript $N$ in the quantities $c$ and $f$ for 
the sake of notational simplicity.  
For $n=2$, and $t_1=t_2=t$, Eq. (\ref{a1}) defines the equal-time 
pair correlation function $f_{i,j}(t)\equiv f_{i,j}(t;t)$.  

An approximate closure to the hierarchy equations
can be obtained 
by performing an expansion in terms of correlation functions
and keeping them up to a certain order, neglecting
the higher-order ones. In this so-called Ursell expansion one defines the 
 correlation functions $f$ through the relations 
\begin{mathletters}
\label{urse}
\ba
\label{a20}
&&c_{i,j}(t)\,=\,c_i(t)\,c_j(t)\,+\,f_{i,j}(t), \\  
\label{a21}
c_{i,j,k}(t)=c_i(t)\,c_j(t)\,c_k(t)\,&&
\,+\,c_k(t)\,f_{i,j}(t)\,
+\,c_i(t)\,f_{j,k}(t)\,+\,c_j(t)\,f_{i,k}(t)\,
+\,f_{i,j,k}(t), \\
&&\mbox{etc.} \nn 
\ea
\end{mathletters} 
where the quantities $f_{i,j}$, $f_{i,j,k}$ account for the fluctuations:
\begin{mathletters}
\ba
f_{i,j}(t)\,&=&\,\langle \delta n_i(t)\,n_j(t) \rangle \\
\label{a22}
f_{i,j,k}(t)\,&=&\,\langle \delta n_i(t)\,\delta n_j(t)\,
\delta n_k(t)\rangle, \\
\mbox{etc.} \nn
\ea
\end{mathletters}
Different approximations emerge from Eqs. (\ref{urse}) when terms 
containing correlation functions of higher orders are neglected. 

In the zeroth approximation all correlations between occupation numbers are
neglected and the state of the system is completely specified by the mean 
occupation numbers $c_i(t)$. From Eq. (\ref{avevol}) we have the MF equation 
\be
\label{a24}
c_i(t+\Delta t)\,=\,c_i(t)\,-\,\frac{1}{N}\,c_i(t)\,
c_{i+1}(t).
\ee
In the first approximation the rate equation (\ref{a24}) is extended 
with a term linear in the NN pair correlation $f_{i,i+1}(t)$,
\be
\label{a25}
c_{i}(t+\Delta t)\,=\,c_i(t)\,-
\,\frac{1}{N}\,c_i(t)\,c_{i+1}(t)\,+
\,\frac{1}{N}\,f_{i,i+1}(t).
\ee
Eq. (\ref{a25}) describes the corrections to the MF equation 
(\ref{a24})
caused by the fluctuations $f_{i,i+1}(t)$ that are built up by sequences of
correlated reactions between the particles. 
These correlations are
calculated from the second hierarchy equation 
(\ref{discor2}) with $j=i+1$. Neglecting 
terms in $o(1/N^2)$ and higher in the r.h.s., we have
\ba
f_{i,i+1}(t+\Delta t)&=&-\frac{1}{N}\,c_{i+1}(t)\,
f_{i,i+2}(t)\,
+\,f_{i,i+1}(t)
\,-\,\frac{1}{N}\,\Big(\,c_i(t)\,c_{i+1}(t)\,-\,
c_i(t)\,[c_{i+1}(t)]^2\big.    \nn \\
\label{a26}
& &\big. +\,c_{i+2}(t)\,f_{i,i+1}(t)\,-
\,c_{i+1}(t)\,f_{i,i+1}(t)\,
+\,f_{i,i+1}(t)\,\Big)
\ea
Here the first term in the brackets represents the propagation to neighboring 
sites of fluctuations correlated over a distance of two lattice spacings. 
The other terms essentially
contain the correlations created or destroyed by the reaction process; 
the terms linear
in $f_{i,i+1}$ represent the effect of existing pre-reaction 
correlations, and the quantity
\be
\label{a27}
\,-\,\frac{1}{N}\,c_i(t)\,c_{i+1}(t)\,+
\,\frac{1}{N}\,c_i(t)\,[c_{i+1}(t)]^2
\ee
accounts for the correlations created by the reaction from a 
completely factorized state. The other pair correlations are calculated
in a similar way from (\ref{discor2}) with $j=i+l$:
\ba
&&f_{i,i+l}(t+\Delta t)=f_l(t)\,-\,
\frac{1}{N}\,c_i(t)\,f_{i+1,i+l}(t) 
\,-\,\frac{1}{N}\,\Big(c_{i+1}(t)+
c_{i+l+1}(t)\Big)\,f_{i,i+l}(t)
\nn \\
\label{a28}
&&\,-\,\frac{1}{N}\,c_{i+l}(t)\,f_{i,i+l+1}(t)+o(1/N^2)
+\mbox{ higher order terms}, \quad
l=2,\ldots,[N]/2.
\ea
In the spatially homogeneous case, the system is translationally 
invariant; the solution obeys the relation $c_i(t)=c(t)$ and the 
correlations $f_{i,j}(t)=f_{|i-j|}(t)$ depend only on the relative distance 
between the evaluation sites.
In the limit $N\to\infty$, Eqs. (\ref{a25}), (\ref{a26}) and (\ref{a28}) 
yield the following set of time-continuous equations
\begin{mathletters}
\label{hifi}
\ba
\label{hi1}
\frac{dc(t)}{dt}&=&-c(t)^2-f_1(t),  \\
\label{hi2}
\frac{df_1(t)}{dt}&=&c(t)^2\,(c(t)-1)-f_1(t)-c(t)f_2(t), \\
\label{hi3}
\frac{df_l(t)}{dt}&=&-c(t)\,[f_{l-1}(t)+2f_l(t)+f_{l+1}(t)], \qquad l= 2,3,
\ldots
\ea
\end{mathletters}
Eqs. (\ref{hifi}) can be used as a starting point for a 
truncation based on the neglect of pair correlations $f_l(t)$ for all $l$
equal or larger than a certain cutoff integer $l_c$. For $l_c=2$, i.e., 
in the pair approximation, the system of two equations approaches the 
steady state $(0,0)$, although the asymptotic
decay of $c(t)$ is much slower than the $t^{-1}$ decay prescribed 
by the MF approach due to large transient effects in $f_1(t)$ 
(Fig. \ref{exvsmfvshie}). A log-log plot of the numerical solution suggests
an inverse power law behavior of $c(t)$ for long times with an exponent 
less than one (see inset in Fig. \ref{exvsmfvshie}).  
In the pair approximation, the stability of the MF fixed point can be 
proven by geometric arguments \footnote{This is illustrated by an analysis 
 of the 2D velocity field}. Numerical results suggest that this property 
holds for any arbitrary
$l_c$, meaning that the system truncated to any order will always attain the  
MF steady state of zero concentration, in contradiction with the non-MF 
results reported above. This reflects the non-negligibility
of the three-point fluctuation correlations $f_{i,j,k}(t)$  
which, as we shall see presently, indeed become of the same order of 
magnitude as the pair correlations $f_l(t)$ already for small $l$.  

One can compute exactly these correlation fluctuations by 
considering the evolution equation for the joint probabilities 
$P_{j,k}^{(l)}(t)$, 
which in our model correspond to the distribution functions
\be
\la \prod_{r=i}^{i+j-1} n_r(t)\,\prod_{s=i+j+l-1}^{i+j+l+k-2} n_s(t) \ra.
\ee
Taking Eq. (\ref{prodev}) as a starting point, a tedious but straightforward 
calculation shows that $P_{j,k}^{(l)}(t)$ evolves according to 
\be
\label{holpriv}
\frac{d P_{j,k}^{(l)}(t)}{dt}=-(j-1)P_{j,k}^{(l)}(t)-(k-1)P_{j,k}^{(l)}(t)
-P_{j+1,k}^{(l-1)}(t)-P_{j,k+1}^{(l)}(t), \quad j,k,l= 1,2,\ldots  
\ee
The first (second) term on the r.h.s. of (\ref{holpriv}) corresponds to 
an event in which a particle inside the cluster of $j\,(k)$ 
occupied sites 
desorbs by interaction with a neighbour inside the cluster, while the third 
(fourth) term stands for the desorption of the rightmost particle
inside the cluster by reaction with an occupied neighbour
site at its right. Eq. (\ref{holpriv}) is solved using the ansatz 
\be
P_{j,k}^{(l)}(t)=p^{j+k-2}\exp[-(j+k-2)t]\,P_{1,1}^{(l)}(t),
\qquad j,k= 1,2,\ldots
\ee
This leads to the (infinite) closed coupled set of equations 
\be
\label{p11}
\frac{d P_{1,1}^{(l)}(t)}{dt}=-p\exp(-t)\left[P_{1,1}^{(l)}(t)+
P_{1,1}^{(l-1)}(t)\right],\qquad l= 2,3,\ldots
\ee
with the boundary condition $P_{1,1}^{(1)}(t)=P_2(t)$.
An explicit expression for $P_{1,1}^{(l)}(t)$ can be obtained from
Eq. (\ref{p11}) by means of a generating function \cite{Jur}.
The final result reads:
\be
P_{1,1}^{(l)}(t)=P_1(t)\left[p\sum_{k=0}^{l-1}
\left[\ln\Bigg(\frac{P_1(t)}{p}\Bigg)\right]^k/k!
+\left[\ln\Bigg(\frac{P_1(t)}{p}\Bigg)\right]^l/l!\right] 
\ee
This allows to compute the fluctuation correlation $f_l(t)$, which is
given by 
\be
\label{cor2p}
f_l(t)=P_{1,1}^{(l)}(t)-[P_1(t)]^2, \qquad l=1,2,\ldots
\ee
This function decreases superexponentially with 
increasing distance $l$, approaching zero for $l \to \infty$.
The three-point NN fluctuation correlation 
$h(t)=f_{1,2}(t)=
\langle \,\delta n_i\,\delta n_{i+1}\,\delta n_{i+2} \,\rangle $ 
can be expressed as 
\be
\label{cor3p}
h(t)=P_3(t)-\left[P_1(t)\right]^3-(2f_1(t)-f_2(t))P_1(t).
\ee
Formulae (\ref{cor2p}) and (\ref{cor3p}) are confirmed 
by numerical simulations (Fig. \ref{corev}). These also show that 
$h(t)$ may already 
get larger than $f_l(t)$ for $l=2$. In fact, the asymptotic values 
$h(\infty)$ and $f_2(\infty)$ are of the same order of magnitude 
in the whole range of the initial coverage $p$ (Fig. \ref{hvsf2}). 
Higher order fluctuations become comparatively large in
the dilute limit $p \ll 1$. 

In the light of the above, the failure of the truncation scheme 
seems thus to be due to the fact that it
includes pair correlations of arbitrary long range whithout taking 
relevant higher order fluctuations into account from the very 
beginning. Although we have only shown the nonnegligibility of 
correlations up to the third order, some numerical computations suggest 
that, regardless of the initial conditions chosen, it is necessary to take 
into account the whole set of fluctuation correlations to obtain a deviation 
from the MF steady state. 

A more restrictive truncation scheme consists in 
neglecting the effect of large clusters in Eqs. (\ref{cgpriv}) 
 by setting $P_k(t)\equiv 0$ beyond a given size \cite{Fot}. 
This truncation procedure yields a 
smooth expansion of the exact results (\ref{dynpk}) 
and (\ref{asco}) in powers of $p$.
 
To conclude this paragraph, let us note that the explicit knowledge
of the joint probabilities $P_{j,k}^{(l)}(t)$ can be used 
to integrate Eqs. (\ref{hol3})-(\ref{hol4}). The solutions of Eqs. 
(\ref{holeq}) read  
\begin{mathletters}
\ba
S_1(t)&=&1-P_1(t), \\
S_2(t)&=&1-2P_1(t)+P_2(t), \\
S_3(t)&=&1+(-3+p-\frac{1}{2}p^2)\,P_1(t)+2P_2(t)-\frac{1}{2}P_3(t), \\
S_4(t)&=&1+(-4+3p-2p^2+\frac{1}{3}p^3)\,P_1(t)+(3-p+\frac{1}{2}p^2)\,P_2(t) 
-P_3(t)\!+\!\frac{1}{6}P_4(t).   
\ea
\end{mathletters}       
They are shown in Fig. \ref{sev} for the case of an initially full lattice. 
\section{Dynamics of the CTD}
\label{ctd}
The starting point is again a 1D lattice with empty ($S$) and 
occupied sites ($A$). The reaction now proceeds according to the scheme 
depicted in Fig. \ref{tdstep}. At each time step, one site is randomly chosen. 
If the chosen site and its neighbour are occupied, {\it both} 
particles will desorb with probability $k_R$. The classical rate 
equation reads 
\be
\frac{d c(t)}{dt}=-2\,k_R \, c(t)^2,
\ee
which yields a faster decay than (\ref{mfeq}) to a zero concentration 
steady state, but again proportional to $t^{-1}$ for long times. It is 
easily seen that the actual 1D system
has the same steady states that the CPD system, i.e. all configurations
with isolated particles. Once more, a deviation from the MF behavior
is found.  

To set up the corresponding microscopic evolution law, we must take into 
account
that this time a particle at a given site $i$ will also desorb if its 
left neighbour
site $i-1$ is chosen and it is occupied. One has therefore an additional 
contribution in the r.h.s. of the dynamical rule:  
\be
\label{dynrul2}
n_i(t+\Delta t)\!=\!n_i(t)\!-\!\xi_N^{(i)}(t)\,\xi_R(t)\,n_i(t)\,n_{i+1}(t)
-\xi_N^{(i-1)}(t)\,\xi_R(t)\,n_{i-1}(t)\,n_i(t), 
\quad i=1,\ldots, N. 
\ee
Like in the CPD case, we can take Eq. 
(\ref{dynrul2}) as a starting point to derive evolution 
equations for the first- and the second-order moments:
\begin{mathletters}
\label{ctdmom}
\ba
\label{1stom}
\frac{d}{dt}\overline{n_i(t)}&=&-\overline{n_{i-1}(t)\,n_i(t)} 
-\overline{n_i(t)\,n_{i+1}(t)} \\[.2cm]
\label{2ndom}
\frac{d}{dt}\overline{n_i(t)\,n_{i+1}(t)}&=&-\overline{n_{i-1}(t)\,n_i(t)} 
-\overline{n_{i-1}(t)\,n_i(t)\,n_{i+1}(t)}-
\overline{n_i(t)\,n_{i+1}(t)\,n_{i+2}(t)} \\[.2cm]
\label{2ndb}
\frac{d}{dt}\overline{n_i(t)\,n_{i+l}(t)}&=&
-\overline{n_i(t)\,n_{i+1}(t)\,n_{i+l}(t)}
-\overline{n_i(t)\,n_{i+l-1}(t)\,n_{i+l}(t)} \nn \\
&&-\overline{n_i(t)\,n_{i+l}(t)\,n_{i+l+1}(t)}
-\overline{n_{i-1}(t)\,n_i(t)\,n_{i+l}(t)}, \qquad{l= 2,3,\ldots}.  
\ea
\end{mathletters}
A generalization of Eqs. (\ref{1stom}) and (\ref{2ndom}) for 
strings of $k$ consecutive sites leads again to a 
hierarchy for the global probabilities $\ti{P}_k(t)$:
\be
\label{privpone2}
\frac{d\ti{P}_k(t)}{dt}=-(k-1)\ti{P}_k(t)-2\ti{P}_{k+1}(t),
\qquad k=1,2,\ldots,k_{max}.
\ee
Note that the prefactor of the second term is now $2$ due to the 
additional reactive event between the leftmost particle inside the 
cluster and a particle at its left neighbour site. An equation of 
the form (\ref{privpone2})
has been obtained in the context of reactant isolation \cite{Coh} 
and random sequential adsorption (RSA) models \cite{Evans,Fan}.  
In these models, the deposition of a dimer on the lattice is equivalent 
to the desorption of a pair of reactants and empty pairs of sites 
available for deposition correspond to unreacted pairs
of reactants (see ref. \cite{Priv}). 

The solutions of the hierarchy (\ref{privpone2}) for a given initial 
configuration read 
\be
\label{tdgensol}
\ti{P}_k(t)=\exp(-(k-1)t)\sum_{j=0}^{k_{max}-k} 
\frac{(2\exp(-t)-2)^j}{j!}\ti{P}_{k+j}(0), \quad k=1,\ldots,k_{max}. 
\ee
(c.f. ref. \cite{Bart}) from which the mean asymptotic coverage follows
straightforwardly:  
\be
\label{genasco2}
c(\infty)=\ti{P}_1(\infty)=
\lim_{N\to\infty}\frac{\overline{N_A(\infty)}}{N}=
\displaystyle{\sum_{j=0}^{k_{max}-1}}
\,\frac{(-2)^j}{ j!}\, \ti{P}_{j+1}(0).
\ee
This result is again in excellent agreement with MC simulations 
(Fig. \ref{tdcov}(a) ).
The coarse-grained solution of Eqs. (\ref{privpone2}) is formally obtained 
by setting $k_{max}=\infty$ and $\ti{P}_j(0)=p^j$. This yields
\begin{mathletters}
\ba
\label{fdcgpro}
P_1(t)&=&p\, \exp{\,[ 2 p\,(\exp{(-t)}-1)}\,], \\
P_k(t)&=&p^{k-1}\exp\left [-(k-1)t\right] P_1(t),\quad k=2,3,\ldots
\ea 
\end{mathletters}
Thus, the corresponding stationary mean coverage takes the form 
\be
\label{fdasco}
c(\infty)=P_1(\infty)=p\,\exp{(-2p)} 
\ee
Note that, contrary to the CPD case, the $p$-dependence of the asymptotic
coverage is no longer monotonic. Thus, as long as  
$p>0.5$, $c(\infty)$ increases when $p$ is decreased 
(Fig. \ref{tdcov}(b) ).  
The initial situation of a fully occupied lattice ($p=1$) corresponds in 
random dimer deposition to an initially empty lattice. In this particular 
case, a 
famous combinatorial argument by Flory \cite{Flor} predicts 
the value $e^{-2}$ for the asymptotic mean fraction of empty sites (in 
our picture, occupied sites) at jamming, in accordance with the formula 
(\ref{fdasco}).   

For the CTD system, it is also possible to derive a truncation
hierarchy by introducing fluctuation correlations in Eqs. 
(\ref{ctdmom}). Like in the CPD case, the second order hierarchy
yields a MF steady state, in contradiction with the exact solution of 
(\ref{privpone2}). Once more, the reason for the failure is the 
nonnegligibility of the three-point correlations, some of which can be 
computed by solving the equation
\ba
\label{genpriv2}
\frac{d P_{j,k}^{(l)}(t)}{dt}&=&-(j-1)P_{j,k}^{(l)}(t)
-(k-1)P_{j,k}^{(l)}(t)\nn -P_{j+1,k}^{(l-1)}(t) \nn \\
&& -P_{j+1,k}^{(l)}(t)-P_{j,k+1}^{(l)}(t)-P_{j,k+1}^{(l-1)}(t), 
\quad j,k,l= 1,2,\ldots    
\ea
which is also found in models for random dimer deposition  
\cite{Evans2}. As shown in \ref{meco}, the 
correlation functions $P_{j,k}^{(l)}$ can be used to set up explicit 
equations for the dynamics of vacant sites. 

Finally, let us mention that it is possible to rederive the moment equations  
(\ref{ctdmom}) and the hierarchy (\ref{privpone2}) from a ME 
analogous to Eq. (\ref{meq}).  
\section{Summary and outlook}
\label{suol}
We have developed a one-dimensional CA model which 
mimics the mesoscopic dynamics of two cooperative desorption
reactions. An important advantage of this approach is that one directly 
sees the effect of a small modification
of the MC algorithm on the underlying moment equations.
The model allows us to derive a hierarchy of linear equations from 
which the mean particle coverage can 
be computed for both microscopic and coarse-grained initial 
conditions. We have seen that a truncation scheme retaining
only pair fluctuation correlations fails to provide the 
correct behavior of these systems, due to the importance of 
higher order fluctuations. However, an alternative truncation 
based on neglecting the effect of large clusters in the dilute 
limit yields a smooth expansion of the cluster dynamics in 
powers of the initial coverage. In the case of the CTD, we 
have pointed out some analogies with models for dimer deposition. 

The CDs studied here are relevant  
both from a theoretical and from an experimental point of view. 
Despite their simplicity, they exhibit a complex non-MF behavior
characterized by nontrivial memory effects and spontaneous ordering. On 
the other hand, a wide number of physical processes on surfaces
involve cooperative desorption of the products \cite{Bow,Zdha}. 
In particular, several problems in the context of exciton dynamics 
in molecular cristals \cite{Ken2,Kop}, recombination of condensed-gas 
radicals \cite{Jack} and 
bond formation in polymers \cite{Flo2,Coh} can be mapped into the 
CDs. A classical example consists of
a long polymer chain (methyl vynil ketone) formed by immobile radical 
groups which react 
with nearest neighbors so as to form rings when the chain is 
heated \cite{Coh}. Clearly, some groups will be isolated by the reaction 
dynamics
and will remain unreacted. In the CTD model, the unreacted groups can be 
identified with the A particles and the rings correspond to pairs 
of vacancies left upon reaction. The quantity of interest, i.e., the 
fraction of unreacted groups, plays then the role of the lattice 
coverage $\theta$.   

The effect of incorporating the backward reaction step $S+A\rightarrow A+A$ 
to the CPD has been studied in refs. \cite{Sud,Prov1,Prak}. 
In this case, the 
mixing properties of the system are restored and one attains a MF
steady state with equal number of empty and occupied sites. It would 
be desirable to extend the present study to the reversible 
CTD and also study the influence of an additional random particle input 
$S\rightarrow A$ on the CDs.   

Another possible extension of our work consists of increasing the range
of the local interactions by allowing, for instance, interactions with next to 
nearest neighbours. One expects that the system approaches the MF 
dynamics as the number of interacting neighbours increases. It is certainly 
worth to characterize this approach in a more quantitative way. 

The dynamics of the one-dimensional CDs changes significantly in 
the diffusion limited case \cite{Lig,Red,Spou,Lus,Fam}. 
One can account for the mobility of the particles by introducing 
additional diffusion terms in the dynamical rule. For the CPD, 
this has been done in ref. \cite{Yo} for initial conditions of the form
(\ref{inco}). As expected, diffusion yields an (anomalous) decay 
into a zero concentration steady state. The results in ref. \cite{Yo} have 
been compared to an off-lattice solution by ben-Avraham 
{\it et al.} \cite{Ben}. Interestingly, the on-lattice solution 
displays a slower decay of the coverage for early times due 
to the finite propagation velocity of a local concentration 
perturbation on the lattice. A CA approach 
for the diffusion-limited CDs has been developed by 
Privman \cite{Priv2}. In the Privman model, all lattice sites are 
synchronously updated at each time step. An extension of this model 
for the diffusionless CDs studied above is also worth carrying out.    
      
A natural generalization of our calculations is to study the dynamics of the 
CDs on Bethe lattices of arbitrary dimension, for which results for the 
cluster dynamics derived heuristically by Majumdar and Privman 
(see \cite{Priv}) are available. For the 
CPD, we expect to obtain results valid for physical lattices as well
by extending some expansion methods developed in the framework of 
dimer deposition \cite{Evans3}.
   
Finally, one would like to extend the boolean CA  
approach to three-state models accounting for the presence of more than
one species and, more generally, to models displaying complex MF 
behavior like oscillations \cite{Alca,Schu,Prov2} and phase transitions 
\cite{Ziff,Marro}.    
\section{Acknowledgments}
We are indebted to F. Baras, J.P. Boon, H.L. Frisch, 
A. Provata and F. Vikas for 
helpful discussions. This work was supported, in part, by the
Training and Mobility of Researchers program of the European Commission
and by the Interuniversity Attractions Poles program of the 
Belgian Federal Government. 
\begin{appendix}
\section{The symmetric CPD}
\label{sym}

In this symmetric version of the CPD model, a particle at the chosen site 
{\it looks} either at
the left or at the right neighbor site with equal probability and 
desorbs if the chosen neighbor is occupied. In this case, the dynamical 
rule reads:
\be
\label{dynrulsym}
n_i(t+\Delta t)=n_i(t)-\xi_N^{(i)}(t)\,\xi_R(t)\,n_i(t)\,\left\{ \xi_L(t)
\, n_{i-1}(t)+(1-\xi_L(t))\,n_{i+1}(t) \right\}, \,\, i\!=\!1,\ldots, N.
\ee 
The additional decision variable $\xi_L(t)$ is equal to one if the left 
neighbour is chosen and zero otherwise. Clearly, the additional choice 
between left and right may change the global coverage in a realization 
characterized by a given path 
$\{\vec{\xi}_N(0),\ldots,\vec{\xi}_N(t)\}$. In contrast, 
one easily checks that averaging Eq. (\ref{dynrulsym}) 
leads to the balance equations (\ref{privpone}) and (\ref{cgpriv})
valid for the asymmetric system. Clearly, the contributions to the 
 destruction of a cluster by reactive events between two particles inside
the cluster
are the same in both models. In the asymmetric model, a cluster may 
also be destroyed by the choice of its rightmost particle, which 
will react with its right occupied neighbour with probability one. In 
contrast, this event will only take place with probability 
$\frac{1}{2}$ in the symmetric model, but there is an additional such
contribution due to the reaction of the leftmost particle in the cluster 
with its left occupied neighbor. Therefore, both models will lead to 
the same balance equations for the cluster probabilities $\ti{P}_k(t)$
and $P_k(t)$.     
This conclusion is supported by comparison of MC simulations
performed according to the rules (\ref{dynrul}) and (\ref{dynrulsym}). A 
similar argument can be applied to the symmetric CTD.
\section{Dynamics of particle islands}
The definition of a particle cluster introduced in \ref{moeq} 
was nonexclusive, meaning that a cluster of a given size could 
contain smaller clusters. In the following, we will consider a more 
restrictive definition in which only isolated strings of particles are
regarded as distinct clusters. These ``particle islands'' are characterized 
by the nonvanishing product  
\be
(1-n_{i-1}(t))\,\left[\prod_{j=i}^{i+k-1}n_j(t)\right]\,(1-n_{i+k}(t)),
\ee
where $k$ is the size of the island. In a $N$-site periodic lattice,  
the total number of islands of size $k$ is given by
\be
N_{I,k}(t)=\sum_{i=1}^N
(1-n_{i-1}(t))\,\left[\prod_{j=i}^{i+k-1}n_j(t)\right]\,(1-n_{i+k}(t)) , 
\ee
and the total number of islands is 
$N_I(t)=\sum_{k=1}^{N-1}N_{I,k}(t)$ (the largest island can at 
most have $N-1$ sites in the ring). 
With this definition, islands can only be created, never destroyed, even
though its size may decrease in time.  
In order to create a new island at time $t$, an existing one must be split 
up by removal of an internal particle (not at the 
edge of an island).
Therefore, the time evolution of $N_I(t)$ will be given by 
\be
N_I(t+\Delta t)=N_I(t)+\sum_{i=1}^N 
\xi_N^{(i)}(t)\,n_{i-1}(t)\,n_i(t)\,n_{i+1}(t).
\ee
In the limit $t\to\infty$, the iteration of this formula yields the
total number of particles $N_A(\infty)$ in the steady state: 
\be
N_A(\infty)=N_I(\infty)=N_I(0)+\sum_{t=0}^{\infty}
\sum_{i=1}^N \xi_N^{(i)}(t)\,n_{i-1}(t)\,n_i(t)\,n_{i+1}(t).
\ee 
The validity of this formula for the asymptotic coverage induced by a 
single realization can be easily checked numerically. 
 
Following paragraph \ref{meco}, we now 
consider an ensemble of realizations starting from the same initial 
condition. The probability to find an 
island of $k$ consecutive sites can be defined in a similar 
way to $\ti{P}_k(t)$:
\be
\ti{I}_k^{(N)}(t)=\frac{\overline{N}_{I,k}(t)}{N}=
\frac{1}{N} \sum_{i=1}^{N} L^{(i)}_k(t), \qquad k=1,\ldots,N-1.
\ee
with 
\be
L^{(i)}_k(t)=\overline{(1-n_{i-1}(t))
\,\prod_{j=i}^{i+k-1}n_j(t)\,(1-n_{i+k}(t))}.       
\ee
After some algebra, one obtains 
\be
\label{iscl1}
\ti{I}_k(t)=\ti{P}_k(t)-2\ti{P}_{k+1}(t)+\ti{P}_{k+2}(t).
\ee
Thus, once the solution for the set of $\ti{P}_k(t)$ is known, 
it is easy to determine the time evolution for the islands. 
A similar relation is found for coarse-grained initial conditions. In 
the continuous-time limit, one has: 
\ba
\label{exclpr}
I_k(t)&&=\langle\, (1-n_{i-1}(t))\,\prod_{j=i}^{i+k-1} n_j(t)\, 
(1-n_{i+k}(t))\,\rangle \nn \\
&&=P_k(t)-2P_{k+1}(t)+P_{k+2}(t)
=[1-2p\,\exp(-t)+p^2\,\exp(-2t)]\,P_k(t). 
\ea
Note that $I_1(t)$ increases monotonically in time, since the 
number of isolated particles always increases. The situation is less
obvious for multiparticle islands (Fig. \ref{probI}). If the initial 
coverage is high enough, their number will first be increased due to the
breaking of larger islands, however they will sooner or later be themselves
reduced to single-particle islands by the ongoing reactions.   
 
The situation is slightly different in the CTD, since two-particle islands 
can indeed be destroyed. The evolution of $N_I(t)$ in a single 
realization is given by:
\ba
N_I(t+\Delta t)&&=N_I(t)+\sum_{i=1}^N 
\xi_N^{(i)}(t)\,n_{i-1}(t)\,n_i(t)
\,n_{i+1}(t)\,n_{i+2}(t) \nn \\
&&-\sum_{i=1}^N 
\xi_N^{(i)}(t)\,(1-n_{i-1}(t)\,)\,n_i(t)\,n_{i+1}(t)\,(1-n_{i+2}(t)\,).
\ea
The first term describes the creation of a new island by 
reaction of two internal particles, while the second term stands for 
the destruction of a two-particle island. The total number of 
particles in the steady state can be written as 
\be
N_A(\infty)=N_I(0)+\sum_{t=0}^{\infty}
\sum_{i=1}^N \xi_N^{(i)}(t)\,n_i(t)\,n_{i+1}(t)\,(n_{i-1}(t)+n_{i+2}(t)-1). 
\ee
The dynamics of the islands can be easily determined by using 
the relations (\ref{iscl1}) and (\ref{exclpr}), which also hold in 
this case.
\end{appendix} 


\newpage 
\begin{figure}[htbp]
\centerline{\psfig{figure=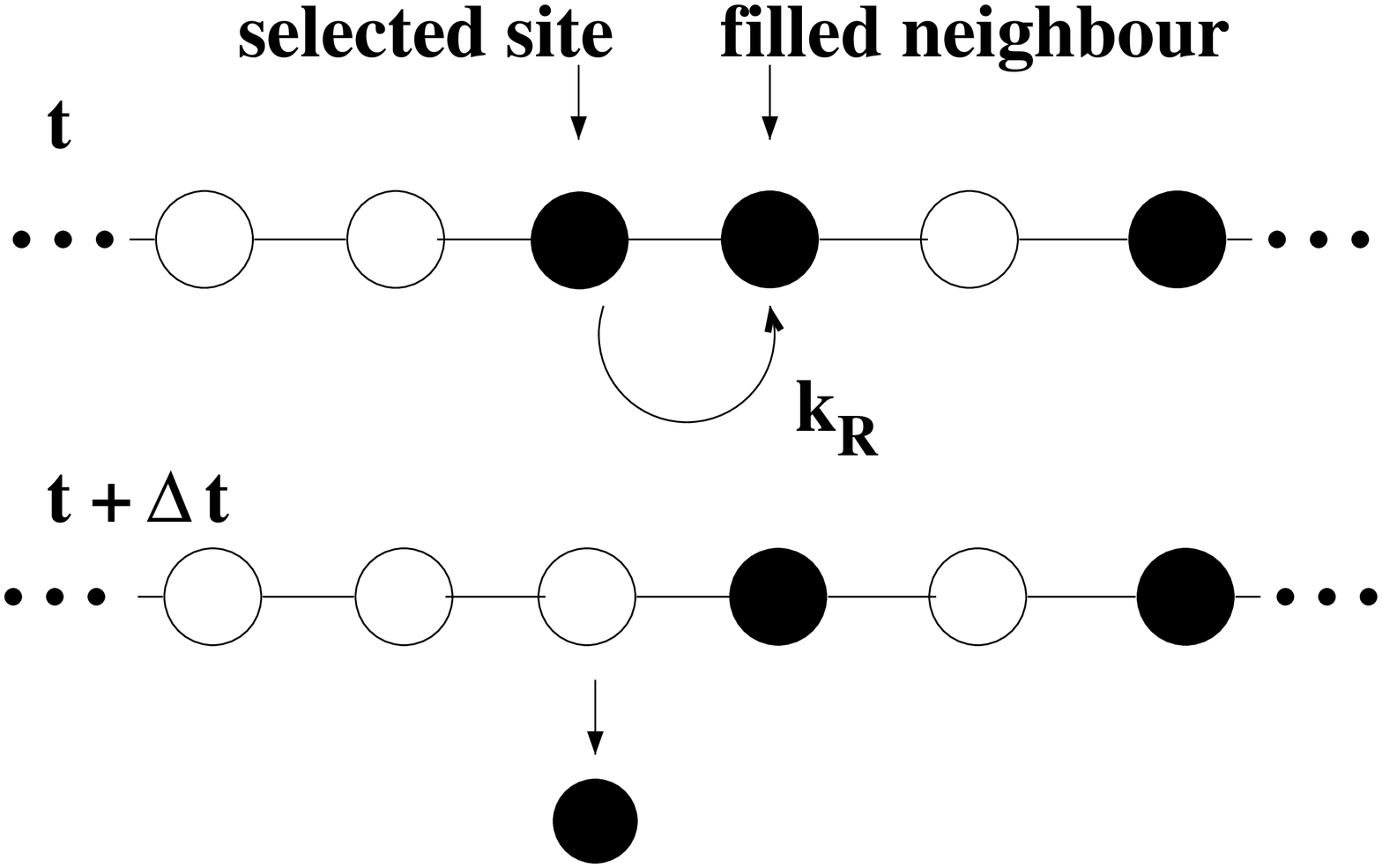,width=7cm,height=5cm}}
\caption{}
\end{figure}
\newpage
\begin{figure}[htbp]
\centerline{\psfig{figure=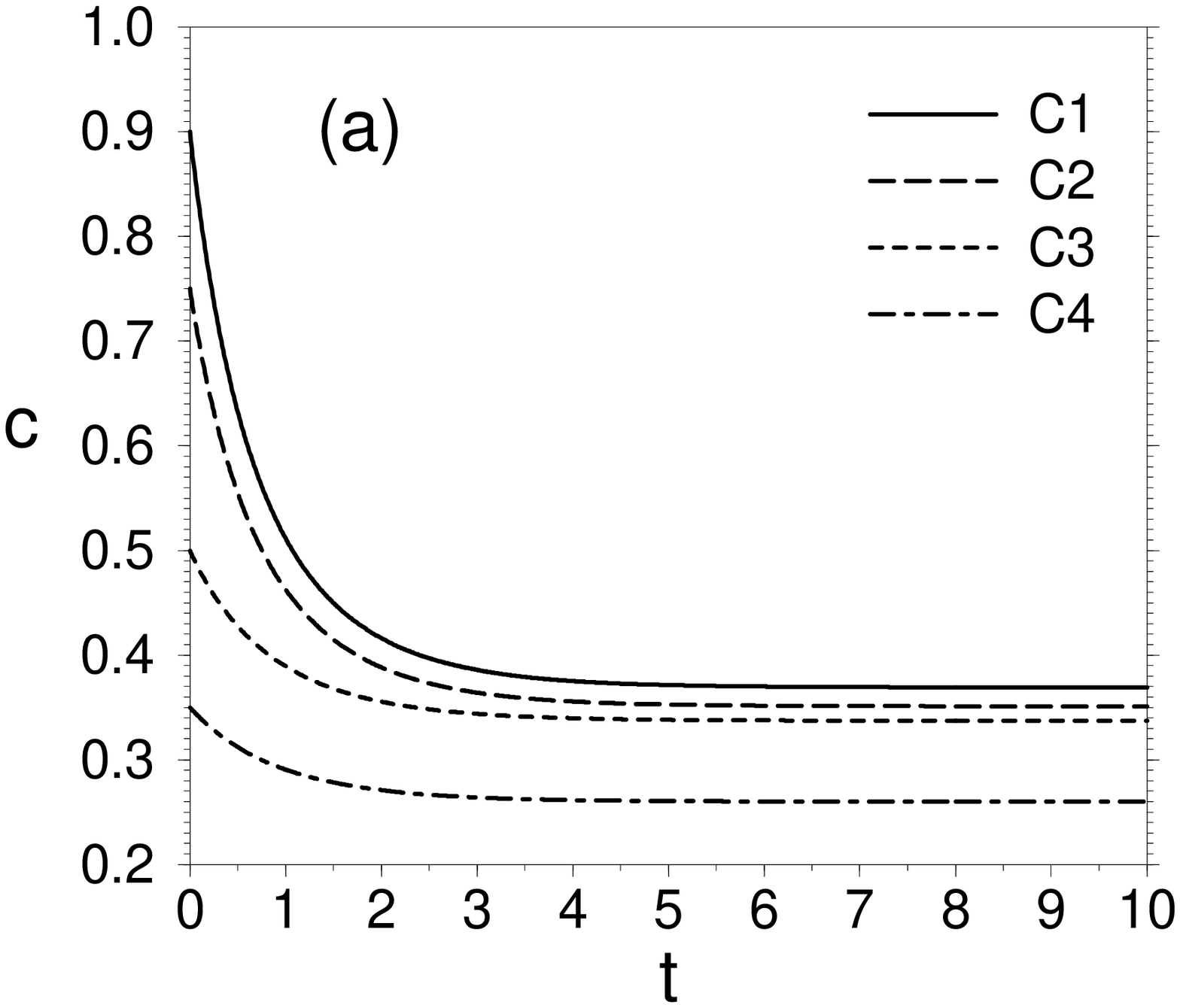,width=10cm,height=8cm}}
\centerline{\psfig{figure=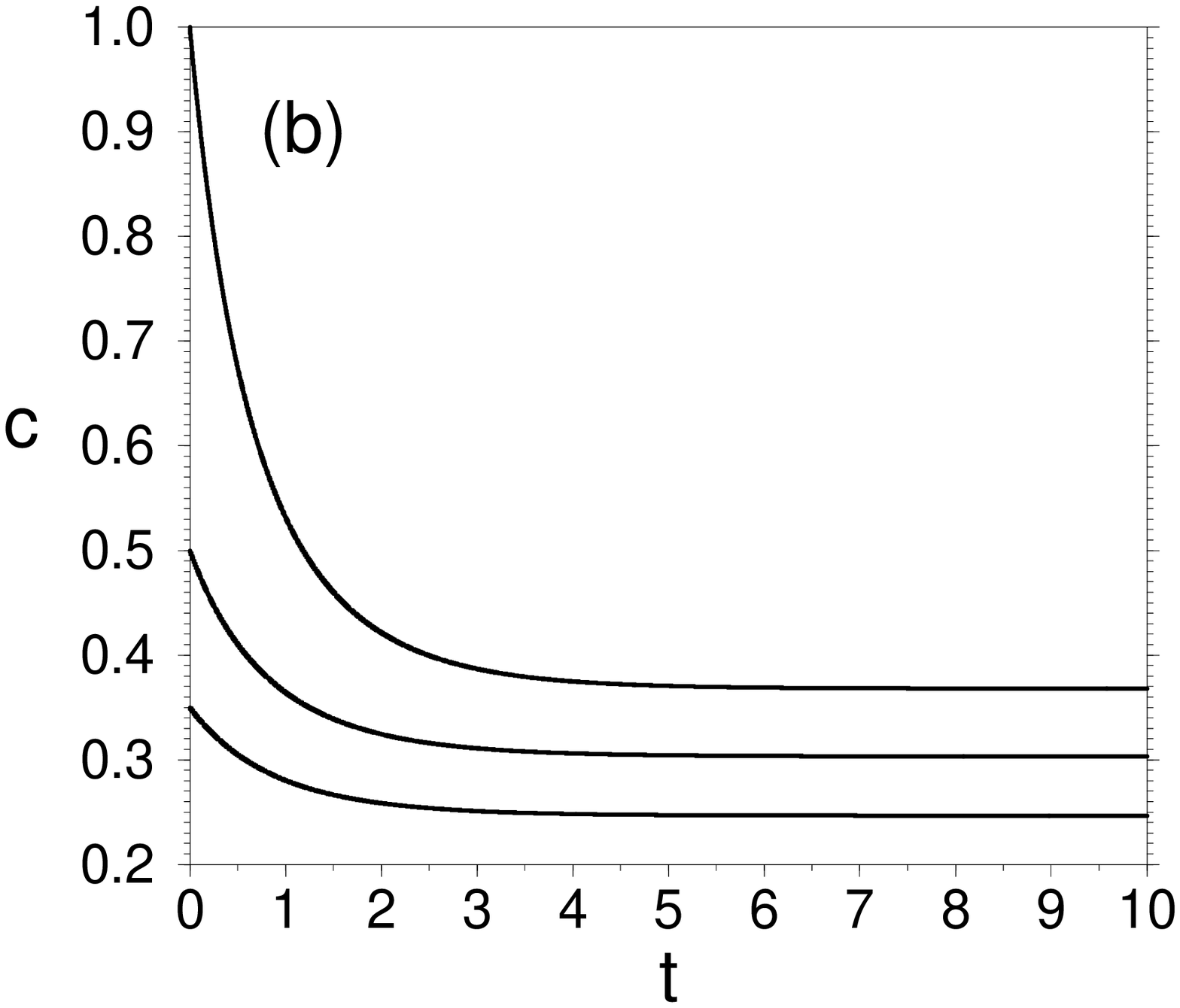,width=10cm,height=8cm}}
\caption{}
\end{figure}
\begin{figure}[htbp]
\centerline{\psfig{figure=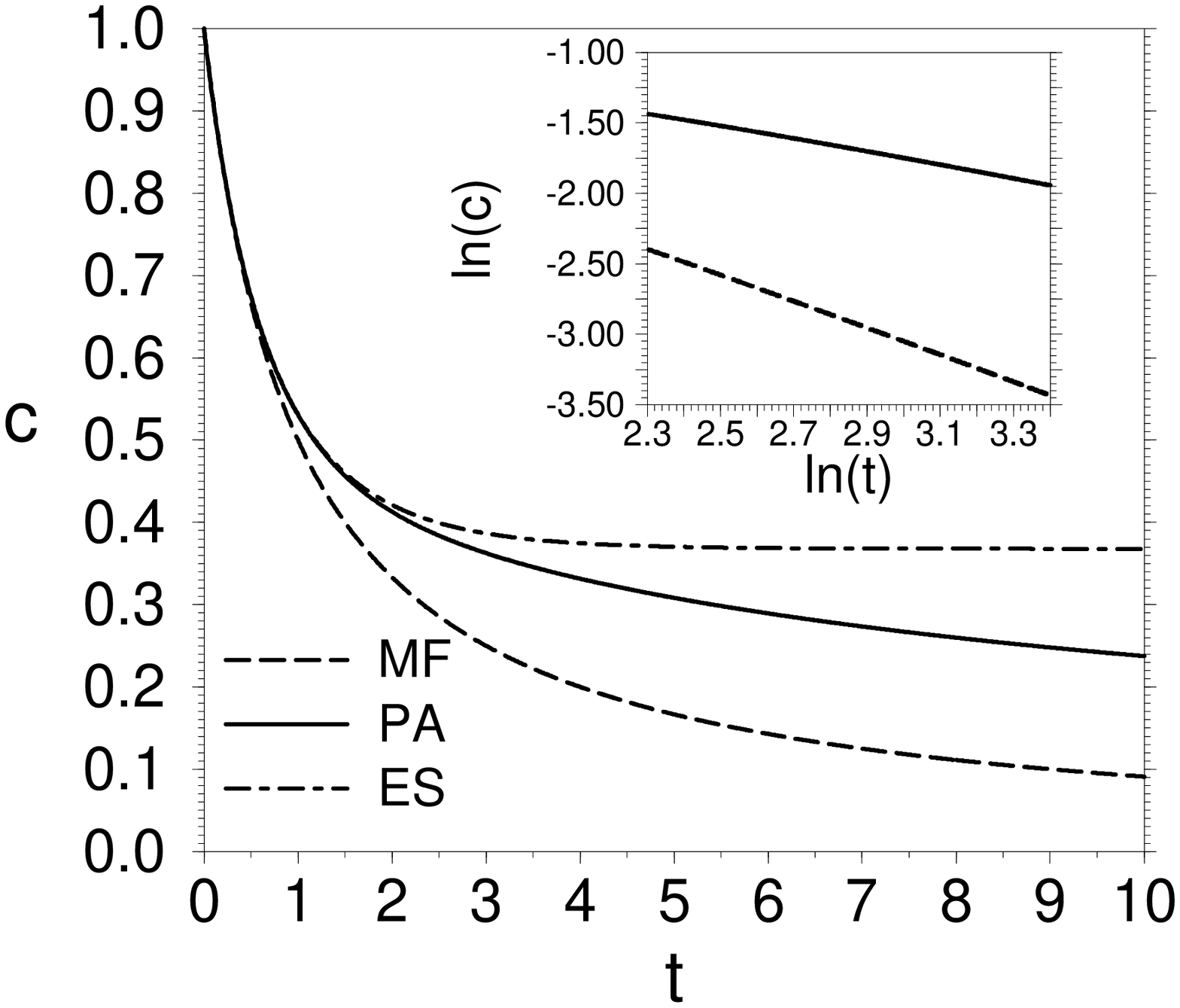,width=10cm,height=8cm}}
\caption{}
\end{figure}
\begin{figure}[htbp]
\centerline{\psfig{figure=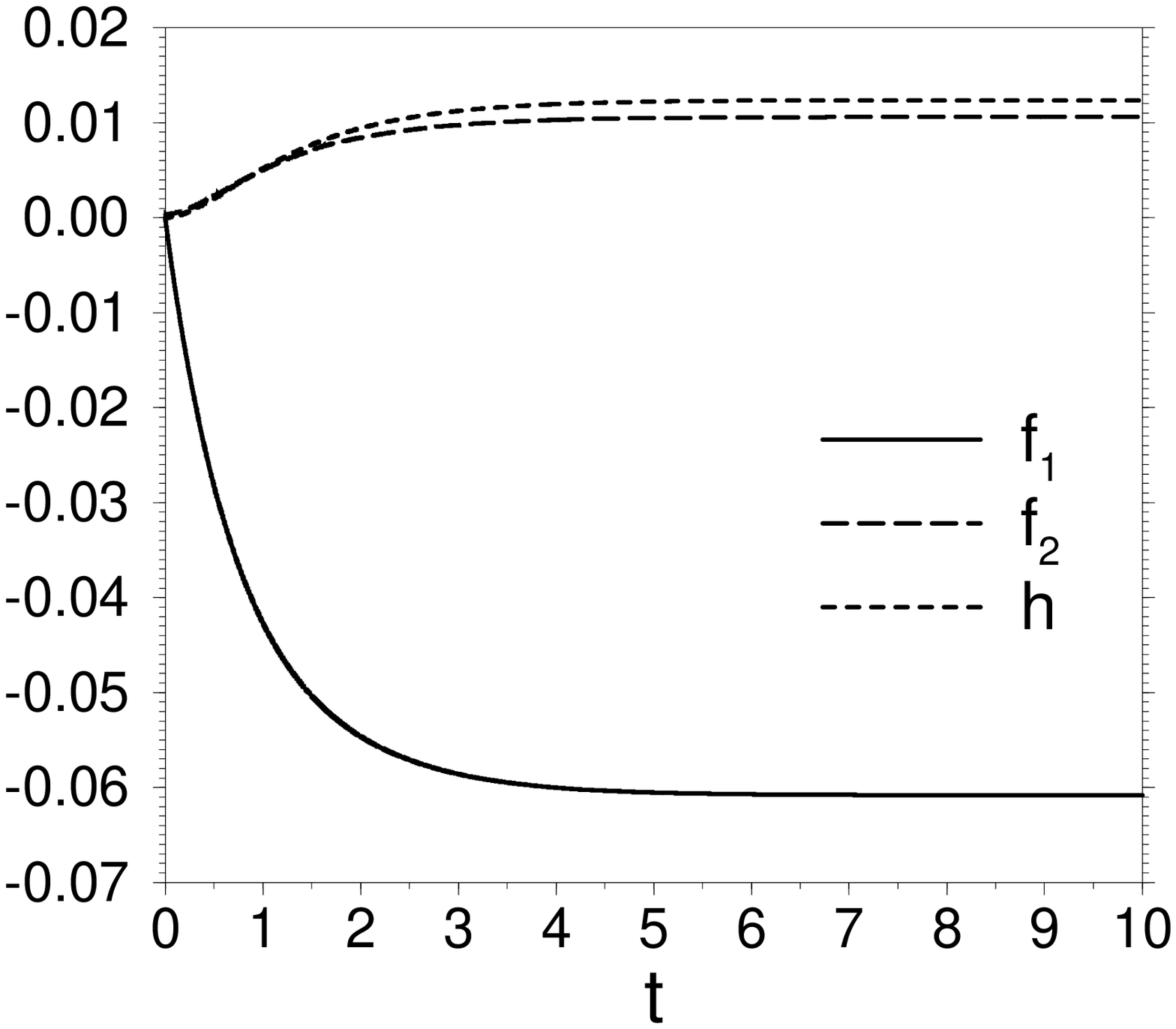,width=10cm,height=8cm}}
\caption{}
\end{figure}
\begin{figure}[htbp]
\centerline{\psfig{figure=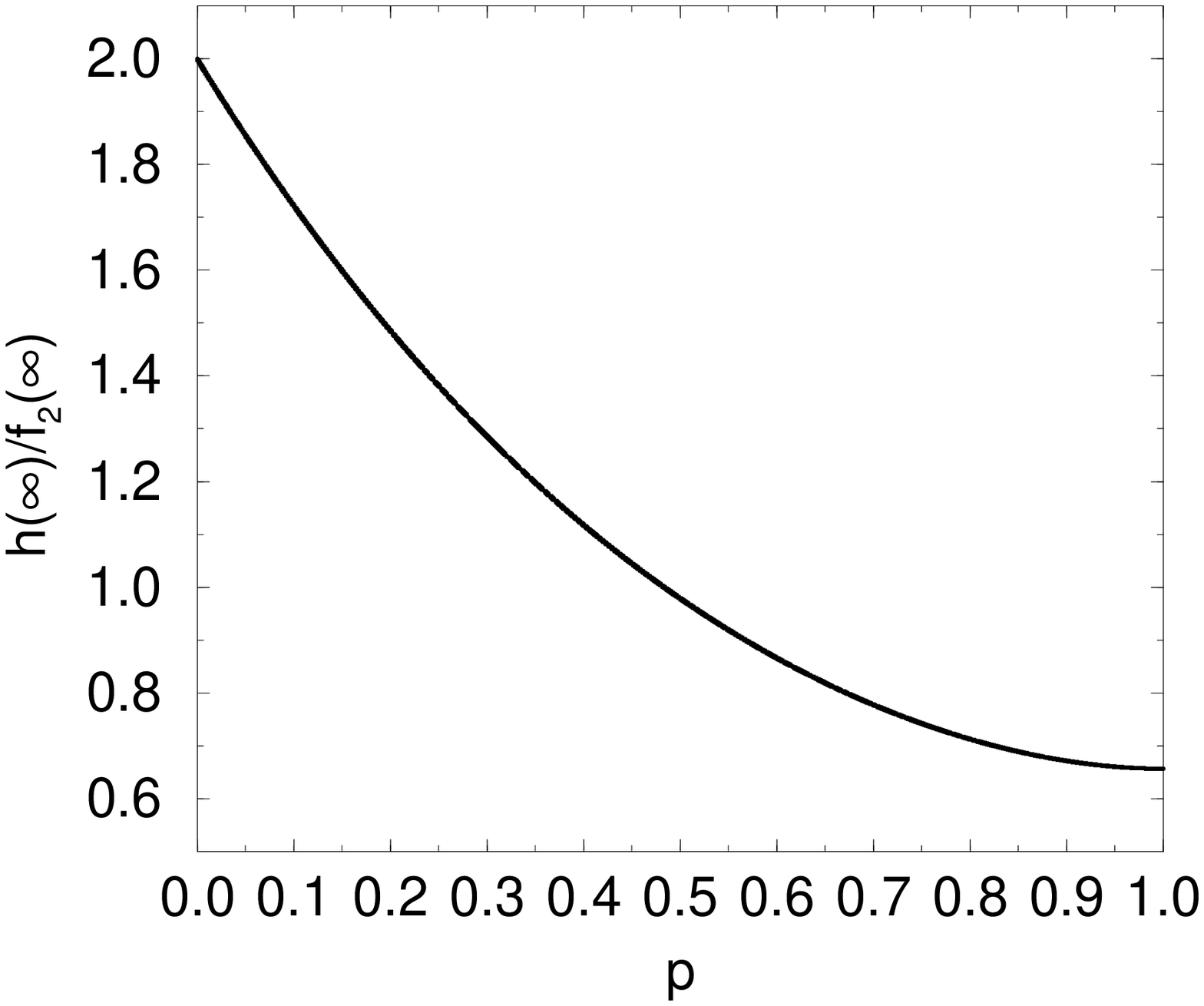,width=10cm,height=8cm}}
\caption{}
\end{figure}
\begin{figure}[htbp]
\centerline{\psfig{figure=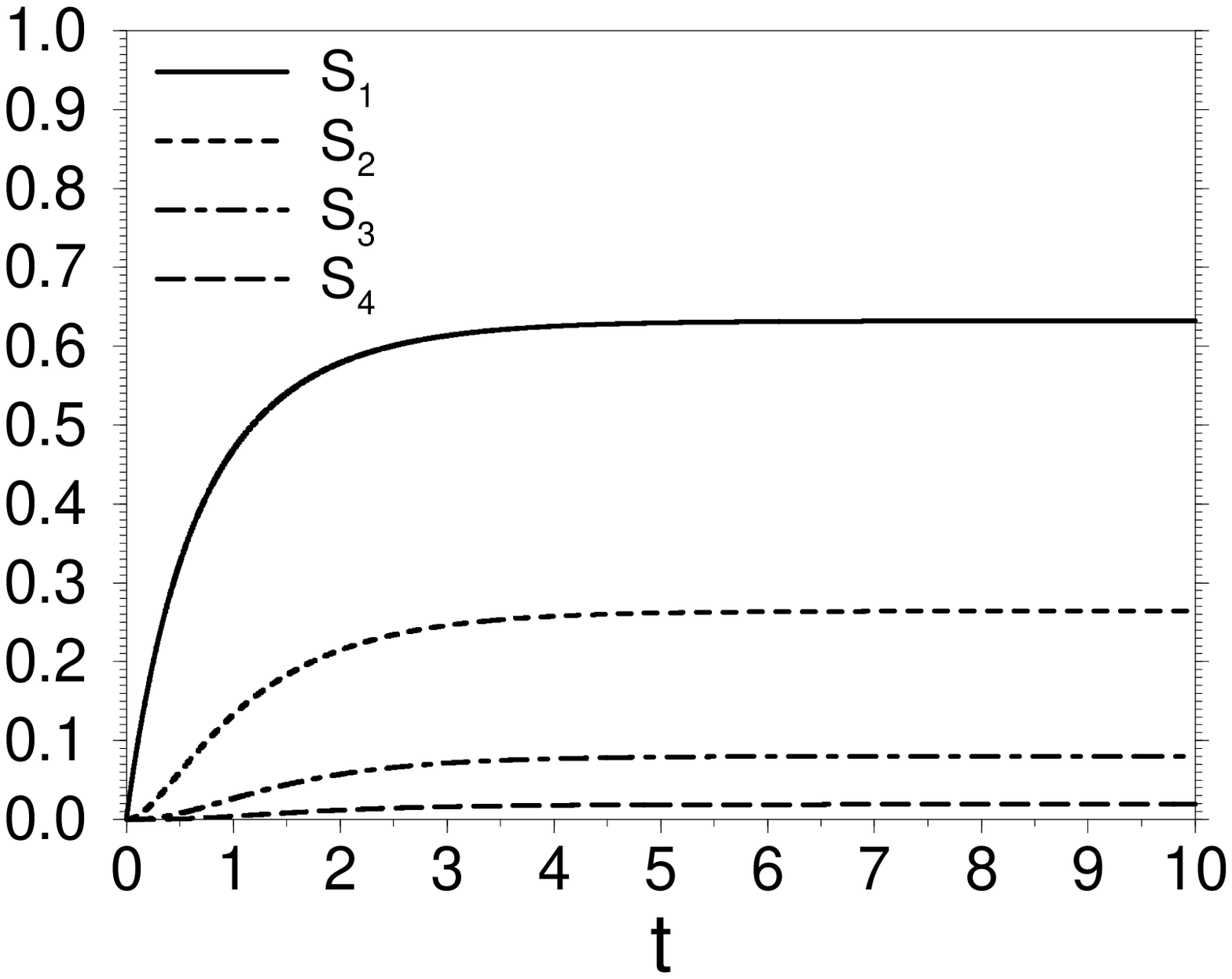,width=10cm,height=7cm}}
\caption{}
\end{figure}
\newpage
\begin{figure}[htbp]
\centerline{\psfig{figure=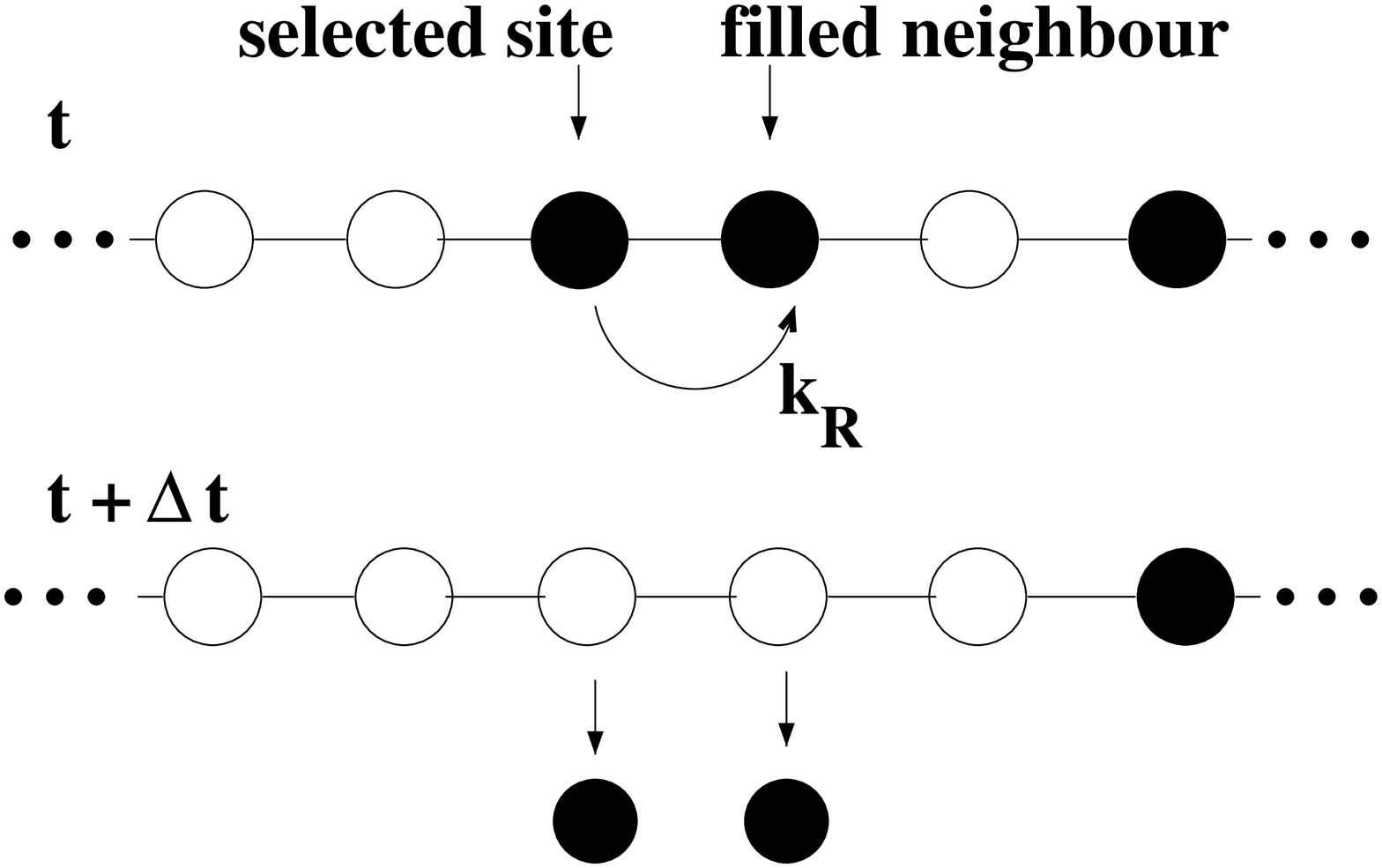,width=7cm,height=5cm}}
\caption{}
\end{figure}
\newpage
\begin{figure}[htbp]
\centerline{\psfig{figure=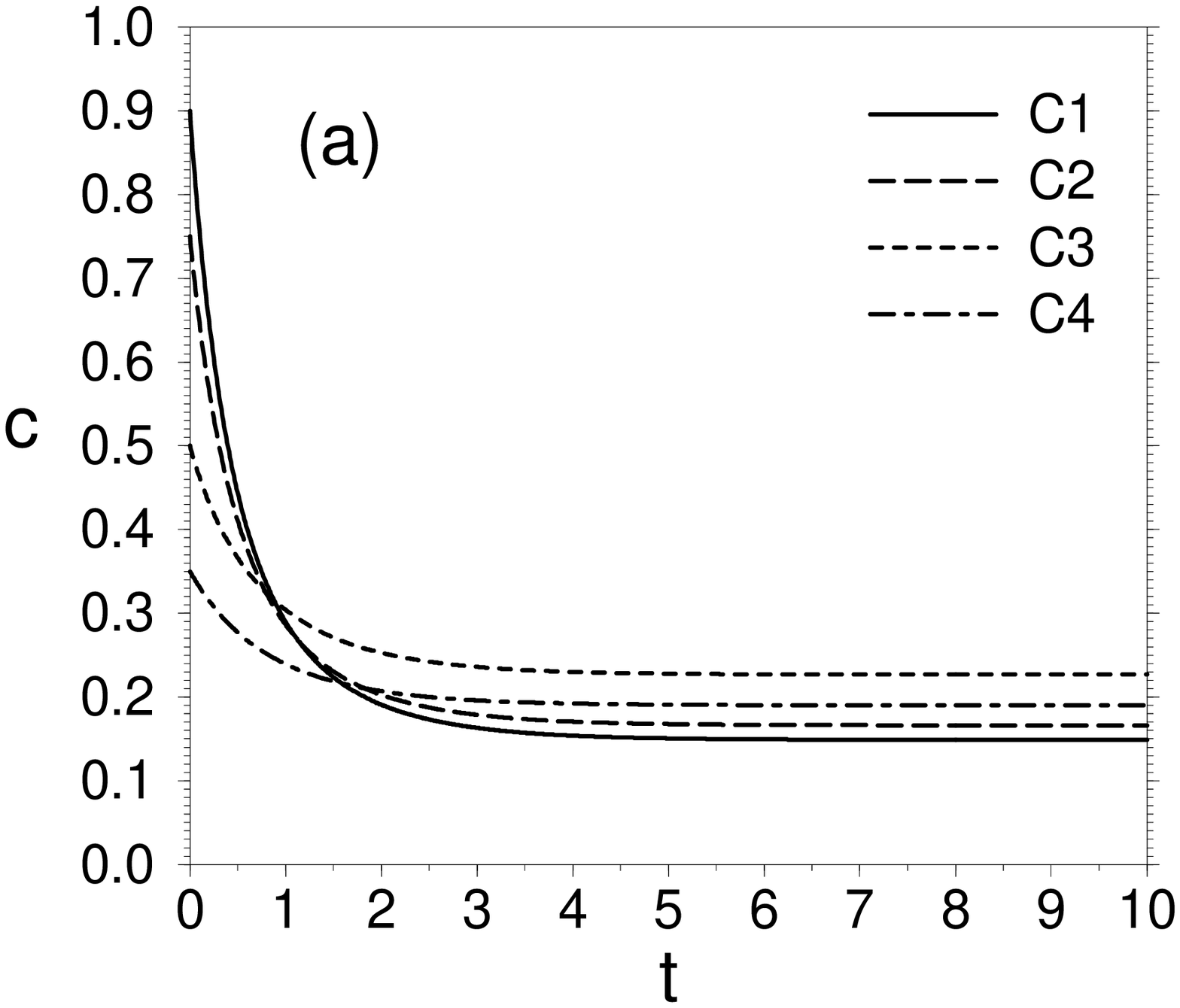,width=10cm,height=8cm}}
\centerline{\psfig{figure=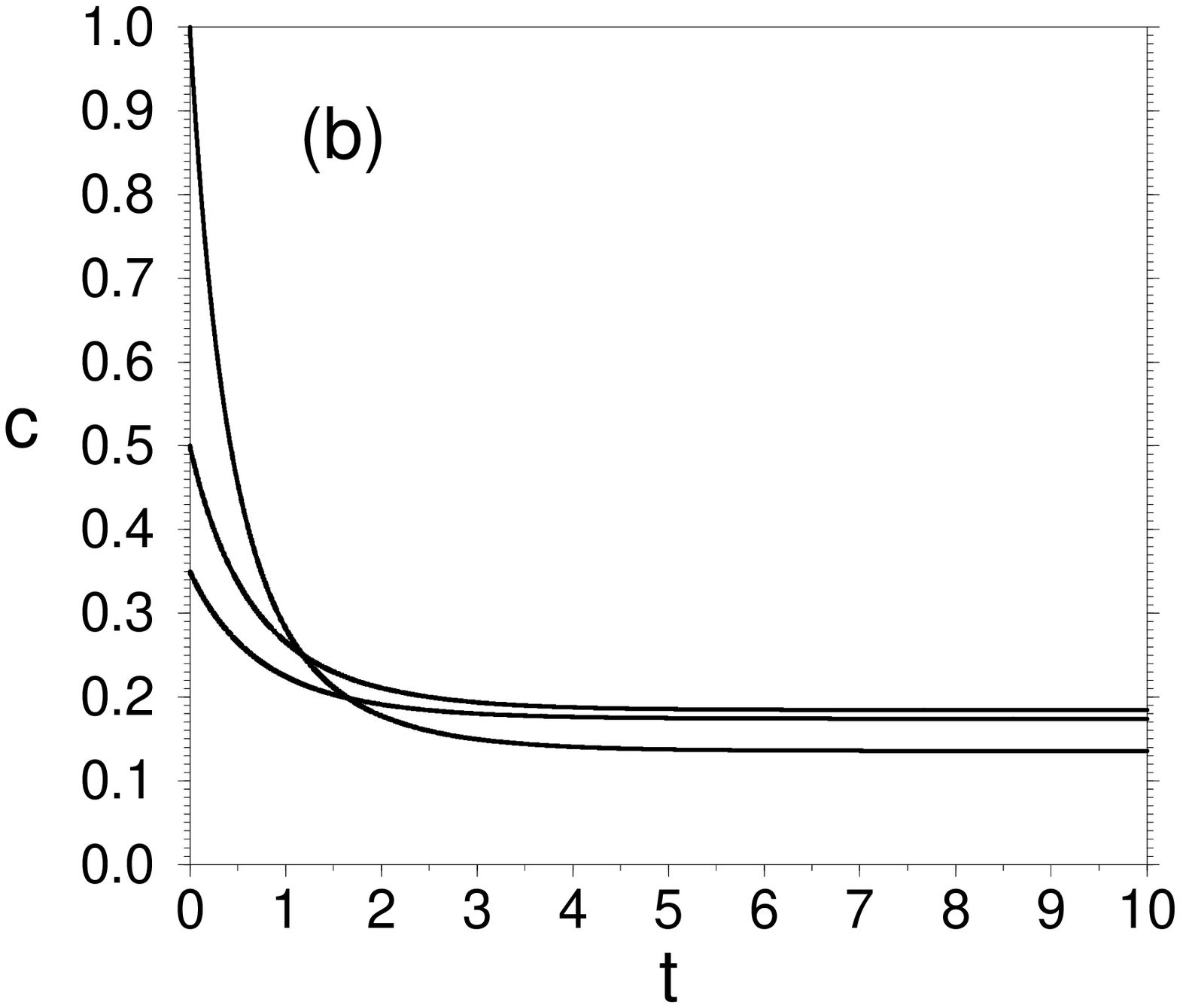,width=10cm,height=8cm}}
\caption{}
\end{figure}
\begin{figure}[htbp]
\centerline{\psfig{figure=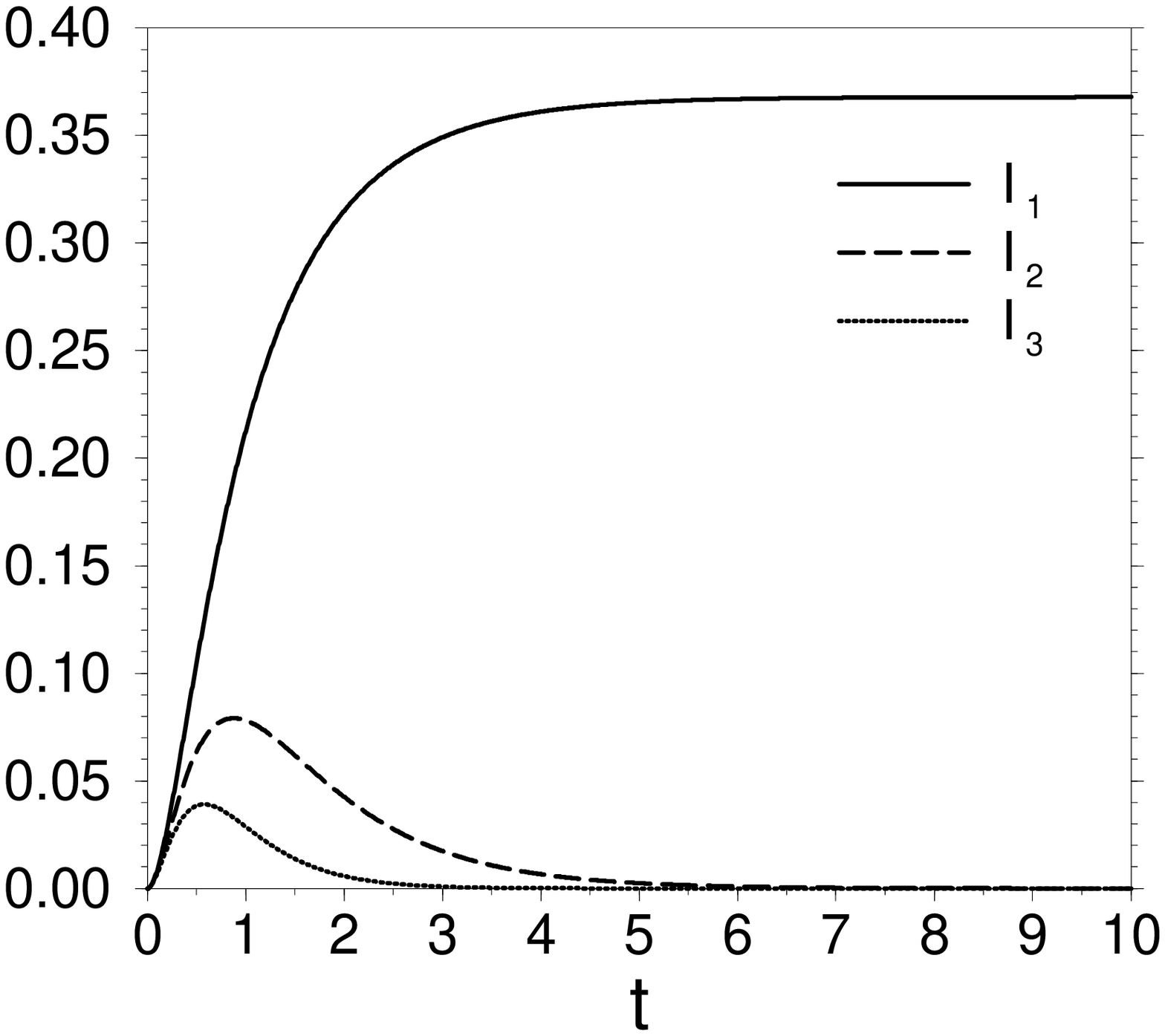,width=10cm,height=8cm}}
\caption{}
\end{figure}
\newpage


\setcounter{figure}{0}

\begin{figure}
\begin{mathletters}
\caption{\label{pdstep} Reaction step for the CPD. If the chosen 
site and a randomly chosen neighbor are filled, the reaction 
proceeds with rate $k_R$. }
\end{mathletters}
\end{figure}

\begin{figure}
\caption{(a) \label{pdcov} Mean coverage $c(t)=\ti{P}_1(t)$ for the four 
different initial configurations $C1-C4$ computed from MC simulations 
over $10^4$ realizations. Both the dynamics and the asymptotic
value of the coverage are in excellent agreement with formulae
(\ref{gensol}) and (\ref{genasco}). The final saturation at different 
coverage values reflects the dependence on the initial conditions, and 
hence, the weak ergodic properties of the system. (b) 
Time evolution of the coarse-grained coverage $c(t)=P_1(t)$ computed from 
MC simulations over $10^2$ realizations for a periodic lattice of 
$10^4$ sites and $p=0.35,0.5$ and $1$ (full lattice). As in Fig. 
\ref{pdcov}(a), the final saturation value depends on the initial 
coverage $p$. Note, in addition, the monotonic dependence for all 
times of $c(t)$ as a function of the initial coverage $p$.}
\end{figure}

\begin{figure}
\caption{\label{exvsmfvshie} Time evolution
of the mean concentration $c(t)$ according to the 
rate equation (\ref{mfeq}) with $k_R=1$ 
(curve MF), the exact solution (\ref{cgpro}) (curve ES) and the numerical 
solution for the pair 
approximation of the truncation hierarchy (\ref{hifi}) (curve PA).
Initially, the lattice is assumed to be fully occupied ($p=1$). In
contrast to the exact solution, the truncated one decays slowly to 
the MF steady state. The inset shows a log-log plot of the concentration
for long times for the MF and the truncated solution. }
\end{figure}

\begin{figure}
\caption{\label{corev} Time evolution of 
$f_1(t), f_2(t)$ and $h(t)$ computed from 
MC-simulations over $10^2$ realizations for a periodic lattice 
with $N=10^4$ and $p=0.35$. Asymptotically, $f_2(t)$ is nearly an order
of magnitude smaller than $f_1(t)$, in accordance with the strong 
spatial decay predicted by formula (\ref{cor2p}). In contrast, the
three-point correlation $h(t)$ is of the same order of magnitude
as $f_2(t)$.}
\end{figure}

\begin{figure}
\caption{\label{hvsf2} $p$-dependence of the asymptotic ratio 
$h(\infty)/f_2(\infty)$
computed from the formulae (\ref{cor2p}) and (\ref{cor3p}). For decreasing
$p$, three-point fluctuations become increasingly important. }
\end{figure}

\begin{figure}
\caption{\label{sev} Analytical solution of the first four evolution 
equations for intervals of vacant sites ($p=1$). The quantities
$S_1$-$S_4$ grow monotonically due to the empty segments created by
the reaction. }
\end{figure}

\begin{figure}
\caption{\label{tdstep} Reaction step for the CTD. In contrast to the 
CPD case, both interacting particles desorb when the event takes place. }
\end{figure}

\begin{figure}
\caption{ \label{tdcov} (a) Mean coverage $c(t)=\ti{P}_1(t)$ for a chain 
ring of $10^2$ sites for the four different initial configurations $C1-C4$  
computed from MC simulations over $10^4$ realizations 
(c.f. Fig. \ref{pdcov}(a) ). Notice, again, the dependence of the asymptotic
coverage on the initial conditions. (b) Time evolution of the coarse-grained 
coverage $c(t)=P_1(t)$ computed from MC simulations over $10^2$ realizations 
for a periodic lattice 
of $10^4$ sites with $p=0.35,0.5$ and $1$ (full lattice). In 
contrast to the CPD case, the behavior of $c(t)$ is no longer monotonic
for sufficiently large times. }
\end{figure}

\begin{figure}
\caption{\label{probI} Time dependence of $I_1(t)$, $I_2(t)$ and $I_3(t)$
according to formula (\ref{exclpr}) for an infinite, initially 
full chain (p=1). }
\end{figure}

\end{document}